\documentclass[aps,prstab,preprint,groupedaddress,showpacs,floatfix]{revtex4}

\usepackage{graphicx}
\usepackage{amssymb}
\usepackage{dcolumn}
\usepackage{bm}

\begin{document}

\preprint{FERMILAB-PUB-09-281-AD-CD}

\title{Fully 3D Multiple Beam Dynamics Processes Simulation for the Tevatron}

\author{E.G. Stern}
\email{egstern@fnal.gov}
\author{J.F. Amundson}
\author{P.G. Spentzouris}
\author{A.A. Valishev}
\affiliation{Fermi National Accelerator Laboratory}

\date{\today}

\begin{abstract}
We present validation and results from
a simulation
of the Fermilab Tevatron including multiple beam dynamics effects.
The essential features of the simulation include a
fully 3D strong-strong beam-beam particle-in-cell Poisson solver,
interactions among multiple bunches and both head-on and long-range beam-beam
collisions,
coupled linear optics and helical trajectory
consistent with beam orbit measurements, chromaticity and
resistive wall
impedance.
We validate individual physical processes against
measured data where possible, and analytic calculations elsewhere.
Finally, we present simulations of the effects of increasing beam intensity
with single and multiple bunches, and study the combined effect of long-range
beam-beam interactions and transverse impedance.
The results of the simulations were successfully used in Tevatron operations
to support a change of chromaticity during the transition to collider mode
optics, leading to a factor of two decrease in proton losses, and thus improved
reliability of collider operations.

\end{abstract}


\maketitle

\section{Motivation}
The Fermilab Tevatron~\cite{Tevatron} is a $p$-$\bar p$ collider operating 
at a center-of-mass energy of~$1.96\,\rm{TeV}$ and peak luminosity reaching 
$3.53\times 10^{32}\,\rm{cm}^{-2}\,~\rm{s}^{-1}$.
The colliding beams consist of 36 bunches moving in a common
vacuum pipe.
For high-energy physics operations, the beams collide head-on at two
interation points (IPs) occupied by particle detectors.
In the intervening arcs the beams are separated by means of electrostatic
separators;
long-range (also referred to as parasitic) collisions 
occur at~136 other locations.
Effects arising from both head-on and long-range beam-beam interactions
impose serious limitations on machine performance, hence constant efforts
are being exerted to better understand the beam dynamics.
Due to the extreme complexity of the problem a numerical simulation appears
to be one of the most reliable ways to study the performance of the system.

Studies of beam-beam interactions in the Tevatron Run II mainly concentrated on
the incoherent effects, which were the major source of particle losses and
emittance growth.
This approach was justified by the fact that the available
antiproton intensity was a factor of~10 to~5 less than the proton intensity
with approximately equal transverse emittances.
Several
simulation codes were developed and used for the optimization of the collider
performance \cite{lifetrac,BBSim}.

With the commissioning of electron cooling in the Recycler, the number of
antiprotons available to the collider substantionally increased. During the
2007 and 2008 runs the initial proton and antiproton intensities differed 
by only a factor of 3.
Moreover, the electron cooling produces much smaller transverse emittance
of the antiproton beam ($\simeq 4 \pi \,\textrm{mm}\,\textrm{mrad}$ 95\% normalized vs. 
$\simeq 20 \pi \, \textrm{mm}\,\textrm{mrad}$ for protons), leading to the head-on beam-beam 
tune shifts of the two beams being essentially equal. 
The maximum attained total beam-beam parameter for protons and antiprotons is 0.028.

Under these circumstances coherent beam-beam effects may become an issue.
A number of theoretical works exist predicting the loss of stability
of coherent dipole oscillations when the ratio of beam-beam parameters
is greater than $\simeq 0.6$ due to the suppression of
Landau damping\cite{Alexahin}. Also, the combined effect of the machine
impedance and beam-beam interactions in extended length bunches
couples longitudinal motion to transverse degrees of freedom and
may produce a dipole or quadrupole mode instability \cite{combinedbb}.

Understanding the interplay between all these effects requires
a comprehensive simulation.
This paper presents a macroparticle
simulation that includes the main features essential
for studying the coherent motion of bunches in a collider: a self-consistent
3D Poisson solver for beam-beam force computation, multiple bunch tracking
with the complete account of sequence and location of long-range and head-on
collision points, and a machine model including our measurement based
understanding of the coupled linear optics, chromaticity, and~impedance.

In Sections \ref{sec:BB3dSec}--\ref{sec:EmitTev} we describe the simulation subcomponents
and their validation against observed effects and analytic calculations.
Section \ref{sec:TevAppSec} shows results from simulation runs which present studies of
increasing the beam intensity.
Finally, in Section \ref{sec:LowerChrom} we study the coherent stability limits
for the case of combined resistive wall impedance and long-range beam-beam interactions.

\section{\label{sec:BB3dSec}BeamBeam3d code}

The Poisson solver in the BeamBeam3d code is described in
Ref.~\cite{Qiang1}.
Two beams are simulated with macroparticles generated with a random
distribution in phase space.
The accelerator ring is conceptually divided into arcs with potential
interaction points at the ends of the arcs.
The optics of each arc is modeled with a $6 \times 6$ linear map that
transforms the phase space $\{x, x', y, y', z, \delta \}$ coordinates of each
macroparticle from one end of the arc to the other.
There is significant coupling between the horizontal and vertical
transverse coordinates in the Tevatron.
For our Tevatron simulations, the maps were calculated using coupled lattice
functions~\cite{Optim1} obtained by fitting a model~\cite{ColOptics}
of beam element
configuration to beam position measurements.
The longitudinal portion of the map 
produces synchrotron motion among the longitudinal coordinates with the
frequency of the synchrotron tune.
Chromaticity results in an additional momentum-dependent phase advance
$\delta \mu_{x(y)} = \mu_0 C_{x(y)} \Delta p/p$
where $C_{x(y)}$ is
the normalized chromaticity for~$x$ (or~$y$) and~$\mu_0$ is the design phase
advance for the arc.
This is a generalization of the definition of chromaticity to apply to
an arc, and reduces to the normalized 
chromaticity $({\Delta \nu}/\nu) / ({\Delta p / p})$ when
the arc encompasses the whole ring.
The additional phase advance is applied to each particle
in the decoupled coordinate basis so
that symplecticity is preserved.

The Tevatron includes electrostatic separators to generate a helical trajectory
for the oppositely charged beams.
The mean beam offset at the IP is included in the Poisson field solver
calculation.

Different particle bunches are individually tracked 
through the accelerator.
They interact with each other with the pattern and locations
that they would have in
the actual collider.

The impedance model applies a momentum kick to the particles generated
by the dipole component of resistive wall
wakefields~\cite{Chao1}.
Each beam bunch is divided longitudinally into slices containing approximately
equal numbers of particles.
As each bunch is transported through an arc, particles in slice~$i$
receive a transverse kick from the wake field induced by the dipole
moment of the particles in forward slice~$j$:
\begin{equation}
{\Delta \vec p_\perp \over p} = {2 \over {  \pi b^3}} 
  \sqrt{ {4 \pi \epsilon_0 c} \over \sigma} \, \,
       {N_j  r_0 \, {< \! \vec r_j \! > } \over {\beta \gamma }} { L \over \sqrt{z_{ij}} }
\label{impedance-kick}
\end{equation}
The length of the arc is~$L$, $N_j$ is the number of particles in
slice~$j$, $r_0$ is the classical electromagnetic radius of the beam particle
$e^2 / 4 \pi \epsilon_0 m_0 c^2$,
$z_{ij}$ is the longitudinal distance between
the particle in slice~$i$ that suffers the wakefield kick and
slice~$j$ that induces the wake. $\vec r_j$  is the mean
transverse position of particles in slice~$j$, $b$ is the pipe radius,
$c$~is the speed of~light, 
$\sigma$~is the conductivity of the beam pipe and $\beta \gamma$ are
Lorentz factors of the beam.
Quantities with units are specified in the
MKSA system.

\section{\label{sec:SBcompSec}Synchrobetatron comparisons}

We will assess the validity of the beam-beam calculation by comparing simulated
synchrobetatron mode tunes with
a measurement performed
at the VEPP-2M
$500\,{\rm MeV}$ 
$e^+ e^-$~collider and described in Ref.~\cite{Nesterenko}.
These modes  are an
unambiguous marker
of beam-beam interactions and provide a sensitive tool
for evaluating calculational models.
These modes arise in a colliding beam accelerator where
the longitudinal bunch length and the transverse beta function 
are of comparable size.
Particles at different~$z$ positions within a bunch
are coupled through the electromagnetic interaction with the
opposing beam leading to the development of
coherent synchrobetatron modes.
The tune shifts for different modes have a characteristic evolution with
beam-beam parameter $\xi = N r_0 / 4 \pi \gamma \epsilon $, in which $N$~is
the number of
particles, $r_0$ is the classical electromagnetic radius of the beam particle, and~$\epsilon$ is
the unnormalized one-sigma beam emittance.

There are two coherent transverse modes in the case of simple beam-beam
collisions between equal intensity beams without
synchrotron motion: the $\sigma$~mode where the two beams oscillate with
the same phase, and the $\pi$~mode where the two beams oscillate with
opposite phases \cite{piwinski}.
Without synchrotron motion, the $\sigma$~mode mode has the same tune
as unperturbed betatron motion while the $\pi$~mode frequency is
offset by $K \xi$, where the parameter $K$ is approximately equal to and
greater than~1
and depends on the transverse shape of the beams~\cite{yokoya}.
The presence of synchrotron motion introduces a more complicated spectrum
of modes whose spectroscopy is outlined in~Fig.~1 in Ref.~\cite{Nesterenko}.

We simulated the VEPP-2M collider using Courant-Snyder uncoupled maps.
The horizontal emittance in the VEPP-2M beam is
much larger than the vertical emittance.
The bunch length ($4\,{\rm cm}$) is comparable
to~$\beta ^*_y = 6\,{\rm cm}$ so we expect to see synchrobetatron modes.
In order to excite synchrobetatron modes, we set an initial $y$~offset
of one beam sigma approximately matching the experimental conditions.

Longitudinal effects of the beam-beam interaction were simulated
by dividing the bunch into six slices.
At the interaction point, bunches drift through each other.
Particles in overlapping slices are subjected to a transverse beam-beam kick
calculated by solving the 2D Poisson equation for the electric field
with the charge density
from particles in the overlapping beam slice.

\begin{figure}[tb]
\includegraphics*[width=\columnwidth]{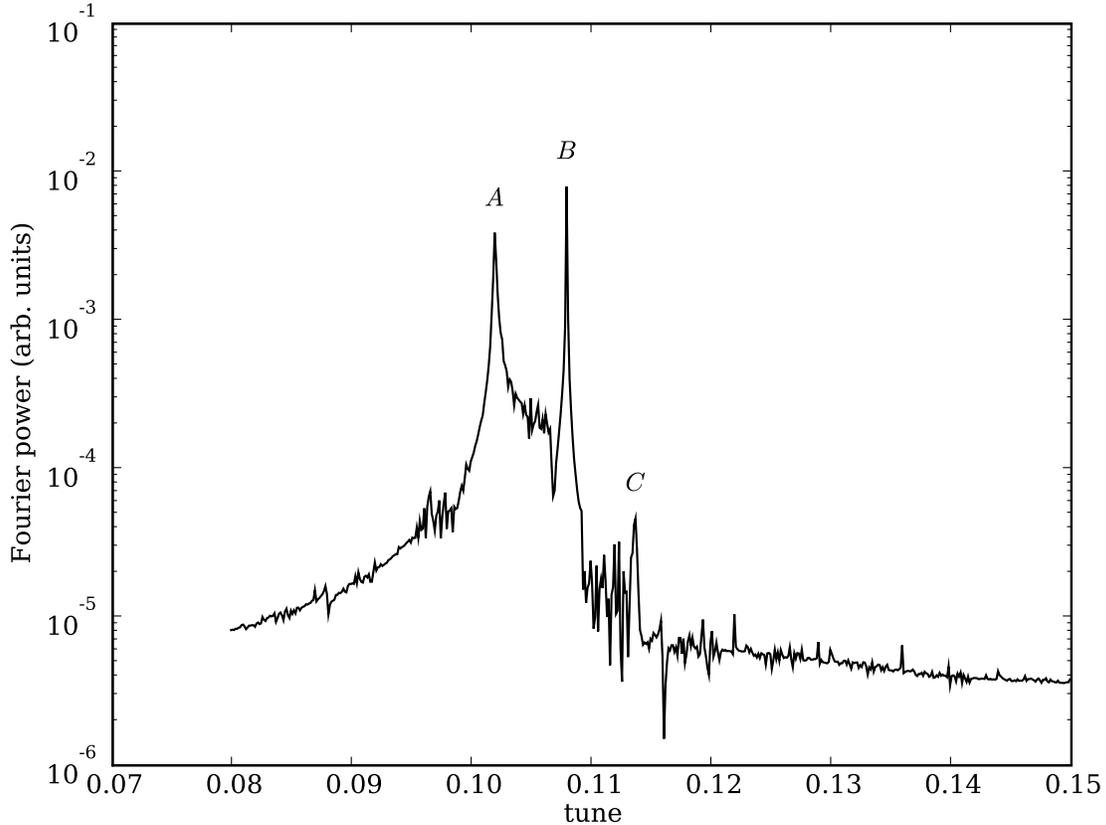}
\caption{\label{sb-modes}Simulated mode spectra in the VEPP-2M collider with
$\xi = .008$ 
showing synchrobetatron modes.
The line indicated by A is the base tune, B is the first synchrobetatron
mode, C is the beam-beam $\pi$ mode. }
\end{figure}

\begin{figure}[tb]
\includegraphics*[width=\columnwidth]{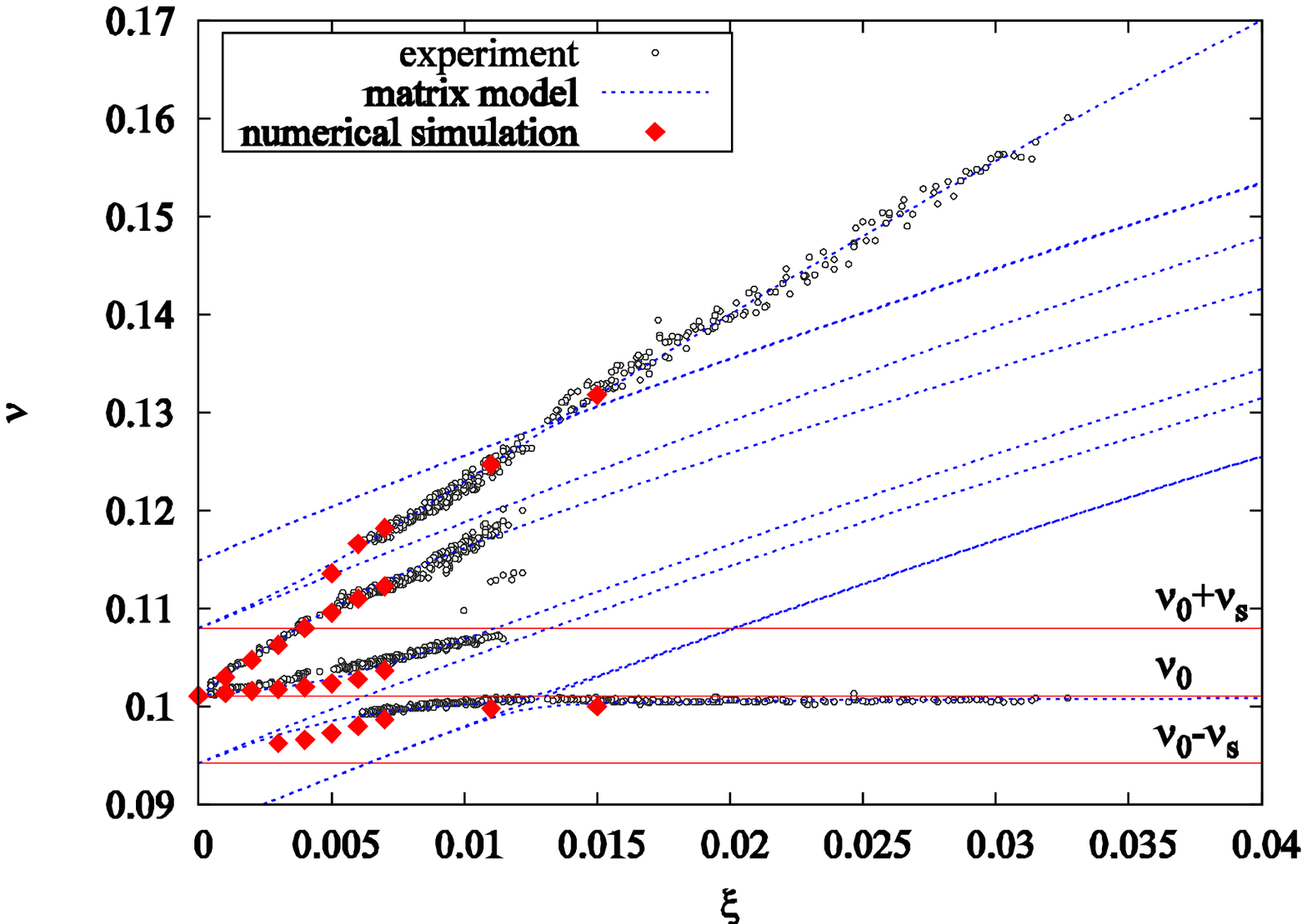}
\caption{\label{vepp-spectra}The diamonds show simulated synchrobetatron modes as a function of
beam-beam parameter $\xi$ (diamonds) and
of observed modes (points).
}
\end{figure}

Simulation runs with a range of beam intensities corresponding to
beam-beam parameters of up to~0.015
were performed, in effect mimicking the experimental procedure described
in~Ref.~\cite{Nesterenko}.
For each simulation run, mode peaks were extracted from the Fourier transform
of the mean bunch vertical position.
An example of the spectrum from such a run is shown in~Fig.~\ref{sb-modes}
with three mode peaks indicated.
In Fig.~\ref{vepp-spectra},
we plot the mode peaks from the
BeamBeam3d simulation as a function of~$\xi$ as red diamonds
overlaid on experimental data from Ref.~\cite{Nesterenko} and a
model using linearized
coupled modes referred to as the matrix model
described in that reference and
Ref.~\cite{Perevedentsev_Valishev, Danilov_Perevedentsev}.
As can be seen, there is good agreement between the observation and simulation
giving us confidence in the beam-beam calculation.

\section{\label{sec:ImpSec}Impedance tests}

Wakefields or, equivalently, impedance in an accelerator with a conducting
vacuum pipe gives rise to well known instabilities.
Our aim in this section is to demonstrate that the wakefield model in
BeamBeam3d quantitatively reproduced these theoretically and
experimentally well understood phenomena.
The strong head-tail instability examined by Chao~\cite{Chao1}
arises in extended length bunches in the presence of wakefields.
For any particular accelerator optical and geometric
parameters, there is an an intensity
threshold above which the beam becomes unstable.

The resistive wall impedance model applies an additional impulse kick in
addition to the application of the map derived from beam optics.
The tune
spectrum is computed from the Fourier transform
of the beam bunch positions sampled at the end of each arc.
In order for the calculation to be a good approximation of the
wakefield effect,
the impedance kick should be much smaller than the~$x'$ or~$y'$
change due to regular beam transport so we divide the ring into multiple
arcs.
 which brings up the question is how many is sufficient.
The difference in calculated impedance tune shift for
a 12~arc division of the ring or a 24~arc division is only~$2\times 10^{-4}$,
which is less than~3\% of the synchrotron tune (0.007 in this study),
the relevant scale in these simulations.
We perform the calculation with~12 arcs for~calculational efficiency.

\begin{figure}[tb]
\includegraphics*[width=\columnwidth]{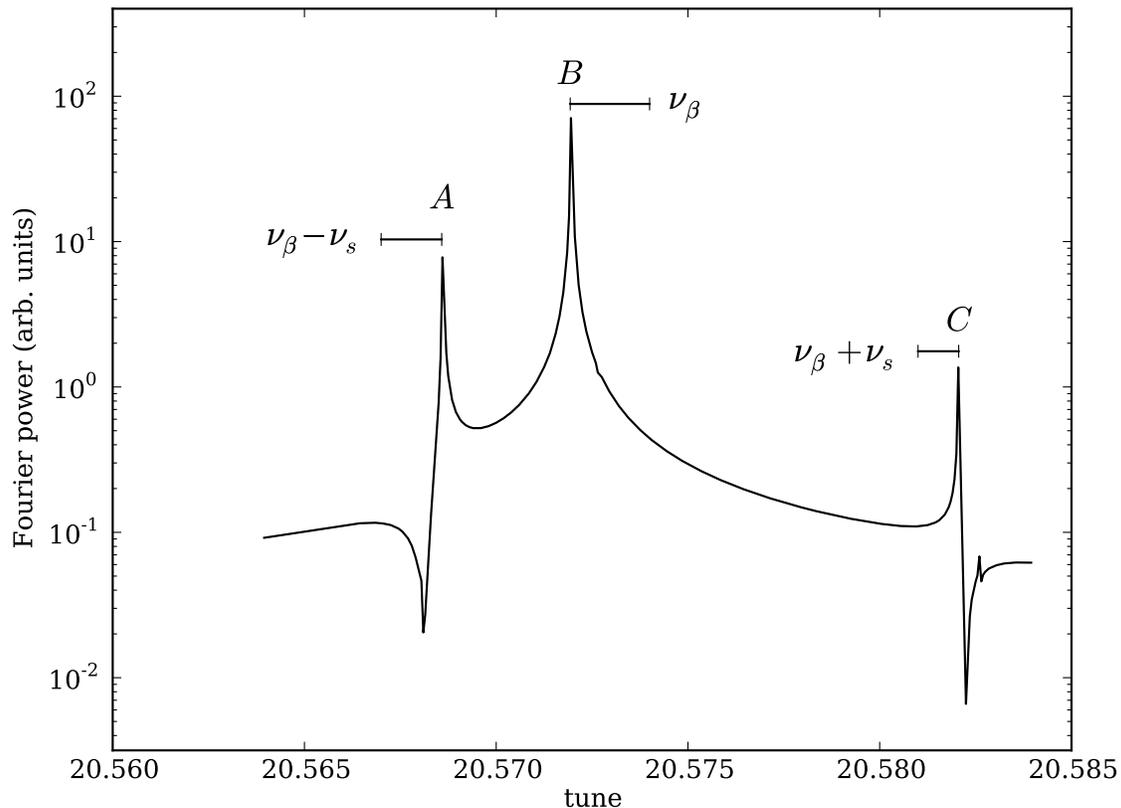}
\caption{\label{impedance-sidebands}Simulated spectrum of a two slice bunch in the presence of
wakefields and synchrotron motion showing three synchrobetatron modes
$A$, $B$, and~$C$ induced by wakefields.
}
\end{figure}

In the absence of impedance, we would expect to see the tune spectrum
peak at~20.574, the betatron tune of the lattice.
With a pipe radius of~$3\,{\rm cm}$ and a bunch length of~$20 \,{\rm cm}$,
resistive wall impedance
produces the spectrum shown in~Fig.~\ref{impedance-sidebands} for a bunch
of~$4\times 10^{12}$ protons
at~$150\,{\rm GeV}$\footnote{We are simulating a much larger intensity than
would be possible in the actual machine in order to drive the strong
headtail instability for comparison with the analytical model.}.
In this simulation, the base tune $\nu_\beta$ is $20.574$ and the synchrotron
tune is~$0.007$.
Three mode peaks are clearly evident corresponding to synchrobetatron modes
with frequencies $\nu_\beta-\nu_s$ shifted up by the wakefield (point~$A$),
$\nu_\beta$ shifted down (point~$B$), and $\nu_\beta+\nu_s$ shifted upward
(point~$C$) as would be expected in Ref.~\cite{chao_strong_headtail}.

\begin{figure}[tb]
\includegraphics*[width=\columnwidth]{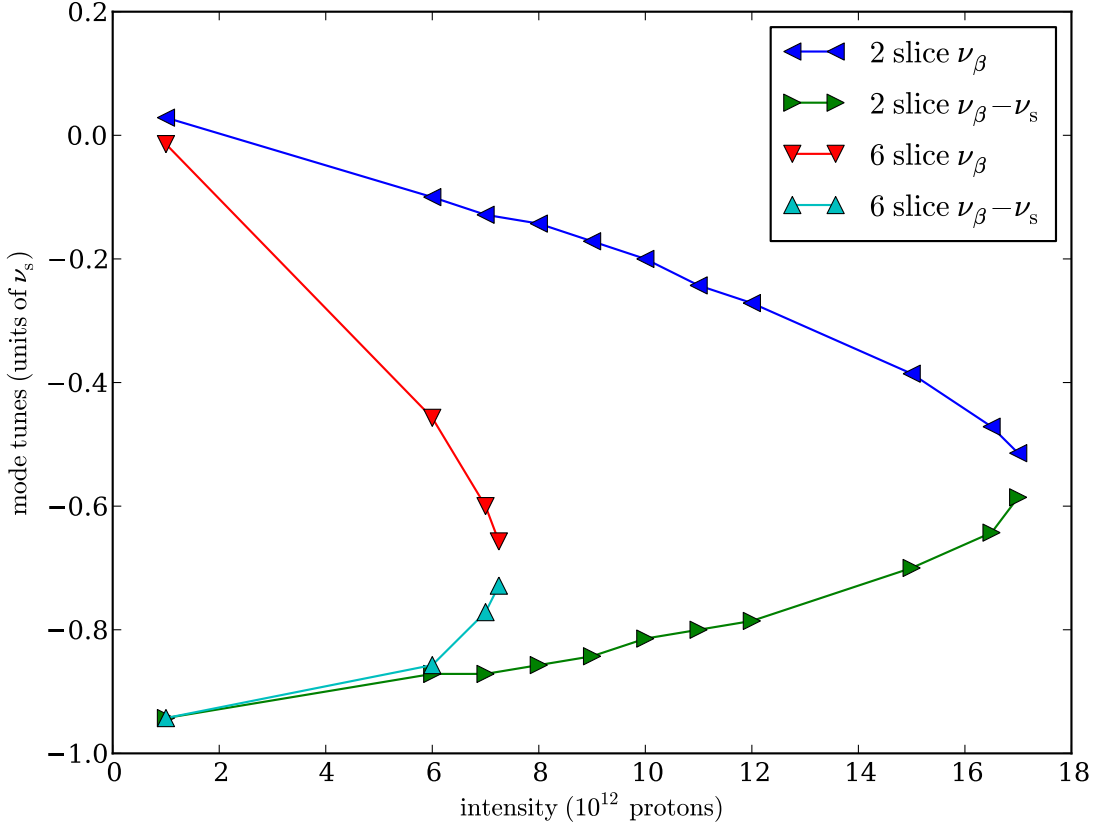}
\caption{\label{strong-ht}
Evolution of the base tune and lower synchrobetatron mode frequencies
as a function of beam intensity showing the two modes approaching a common
frequency due to impedance.
The~$y$ scale is in units of the synchrotron tune.
The simulations are shown for a two slice aond  six
slice wakefield calculation.}
\end{figure}

In Fig.~\ref{strong-ht}, we show the evolution of the two modes as a function
of beam intensity.
With the tune and beam environment parameters
of this simulation, Chao's two particle model  predicts instability
development
at intensities of about~$9 \times 10^{12}$ particles, which is close to
where the upper and lower modes meet.
We show two sets of curves for two slice and six slice wakefield
calculations.
The difference between the two slice and six slice simulations is
accounted for by the effective slice separation, $\hat z$, that enters
Eq.~\ref{impedance-kick}.
With two slices, the effective~$\hat z$ is larger than than the six slice
effective~$\hat z$, resulting in a smaller~$W_0$.
With the smaller wake strength, a larger number of~protons is
necessary to drive the
two modes together as is seen in Fig.~\ref{strong-ht}.

\begin{figure}[tb]
\includegraphics*[width=\columnwidth]{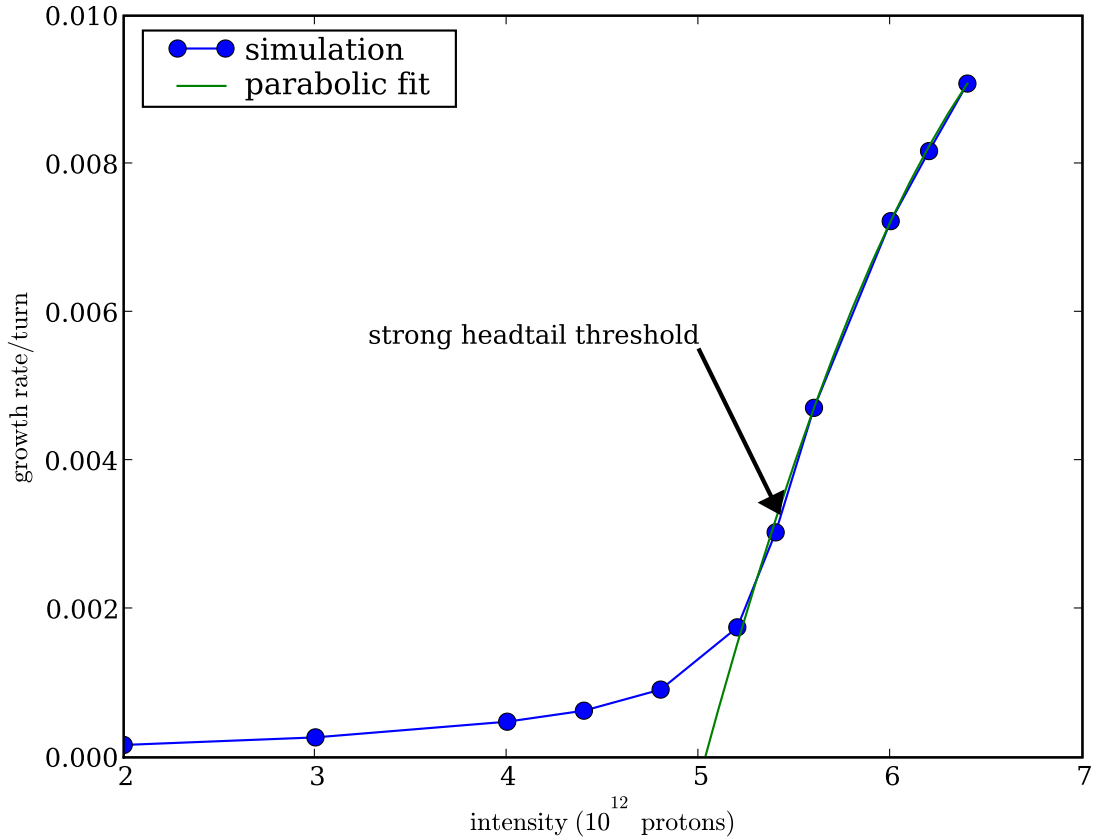}
\caption{\label{strong-ht-thresh}The growth rate
of dipole motion in the simulated accelerator
with impedance
as a function of beam intensity as the strong head-tail threshold is 
reached superimposed with a parabolic fit.}
\end{figure}

When the instability occurs, the maximum excursion of the bunch dipole
moment grows exponentially as the beam executes turns through the
accelerator.
The growth rate can be determined by reading the slope of a graph of the absolute
value of bunch mean position as a function of turn number plotted on
a log scale.
The growth rate per turn of dipole motion at the threshold of strong head-tail
instability
has a parabolic dependence on beam intensity.
The wakefield calculation reproduces this feature, as shown in
Fig.~\ref{strong-ht-thresh}.
The growth rate is slowly increasing up to the instability threshold at
$5.42\times 10^{12}$, after which it has the explicitly quadratic
dependence on beam intensity ($I$) of
$\textrm{growth rate} = -0.100 + 0.0304 I - 0.00207 I^2$.

\begin{figure}[tb]
\includegraphics*[width=\columnwidth]{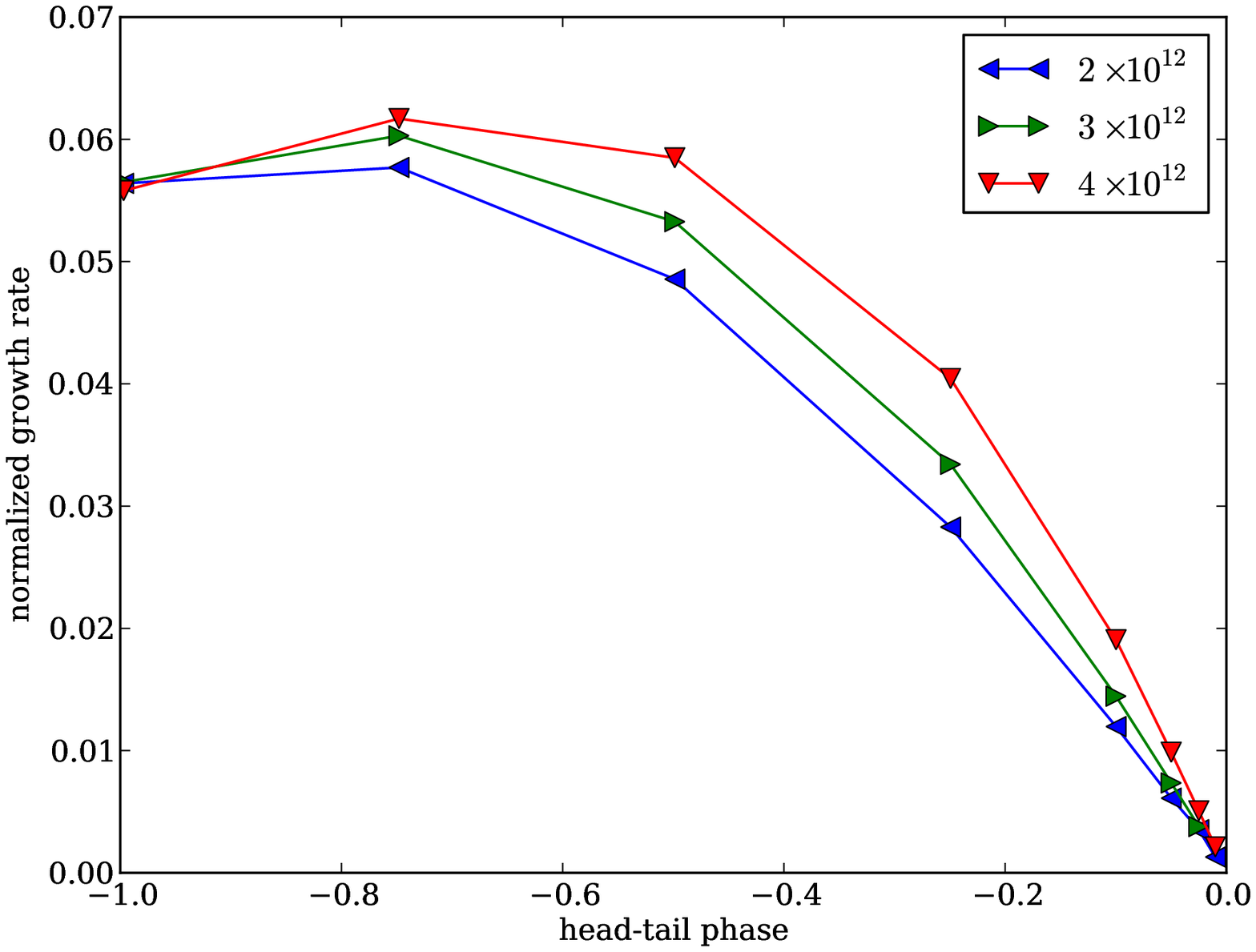}
\caption{\label{norm-growth-vs-htphase}The normalized growth rate
of dipole motion in the simulated accelerator
with impedance and chromaticity
as a function of head-tail phase $\chi$ at three beam intensities demonstrating
their linear relationship close to~$0$ and the
near-universal relationship for
head-tail phase between~$-1$ and~$0$.}
\end{figure}

Chromaticity interacts with impedance to cause
a different head-tail instability.
We simulated a range of beam intensities and chromaticity values.
The two particle model and the more general Vlasov equation
calculation~\cite{Chao1} indicate that the growth rate scales by
the head-tail phase $\chi = 2 \pi C \nu_\beta {\hat z} / c \eta$, where
 $\eta$ is the
slip factor of the machine
and~$\hat z$ is roughly the bunch length.
The head-tail phase gives the size of betatron phase variation
due to chromatic effects over the length of the bunch.

Some discussion of the meaning of the slip factor in the context of a
simulation is necessary.
In a real accelerator, the slip factor has an unambiguous meaning:
$\eta = (\alpha_C - 1/\gamma^2)$.  The momentum compaction parameter $\alpha_c$
is determined by the lattice and $\gamma$ is the Lorentz factor. 
We simulate longitudinal motion by applying maps to the particle
coordinates~$z$  and~$\delta$ in discrete steps.
The simulation parameters specifying longitudinal
transport are the longitudinal beta function~$\beta_z$ and synchrotron
tune $\nu_s$.
Note that these parameters do not make reference to path length travelled
by a particle.
However, path length enters into the impedance calculation because
wake forces are proportional to path length.
In addition, analytic calculations of the effect of wake forces 
depend on the evolution of the longitudinal particle position which in
turn depend explicitly on the slip factor.
For our comparisons with analytic results to be meaningful,
we need to use a slip
factor that is consistent with the longitudinal maps and the path lengths
that enter the wake force calculations.
The relationship between the slip factor~$\eta$ and the simulation
parameters is $\beta_z = \eta L_{\bigcirc} / 2 \pi \nu_s$, where~$L_{\bigcirc}$ is the length
of the accelerator and 
$\beta_z = \sigma_z/\sigma_\delta$ is the longitudinal beta function\cite{Chao2,Syphers}
which may be derived by identifying corresponding terms in the solution to
the differential equations of
longitudinal motion and a one term linear map.

When the growth rate is normalized by~$N r_0 W_0/2 \pi \beta \gamma \nu_\beta$,
which includes the beam intensity and geometric factors, we expect a universal
dependence of normalized growth rate versus head-tail phase that begins linearly
with head-tail phase\cite{Chao3} and peaks around -1\footnote{%
The simulated machine is above transition ($\eta$ is positive.)
The head-tail instability develops when chromaticity is negative, thus
the head-tail phase is negative.
}.

Fig.~\ref{norm-growth-vs-htphase} shows the simulated growth rate 
at three intensities
with a range of chromaticites from $-.001$ to~$-0.5$ to get head-tail phases
in the~$0$ to~$-1$ range.
The normalized curves are nearly identical and peak close to head-tail
phase of~unity.
The deviation from a universal curve is again due to differences between
the idealized model and  detailed simulation.

\section{\label{sec:EmitTev}Bunch-by-bunch emittance growth at the Tevatron}

\begin{figure}[tb]
\includegraphics*[width=\columnwidth]{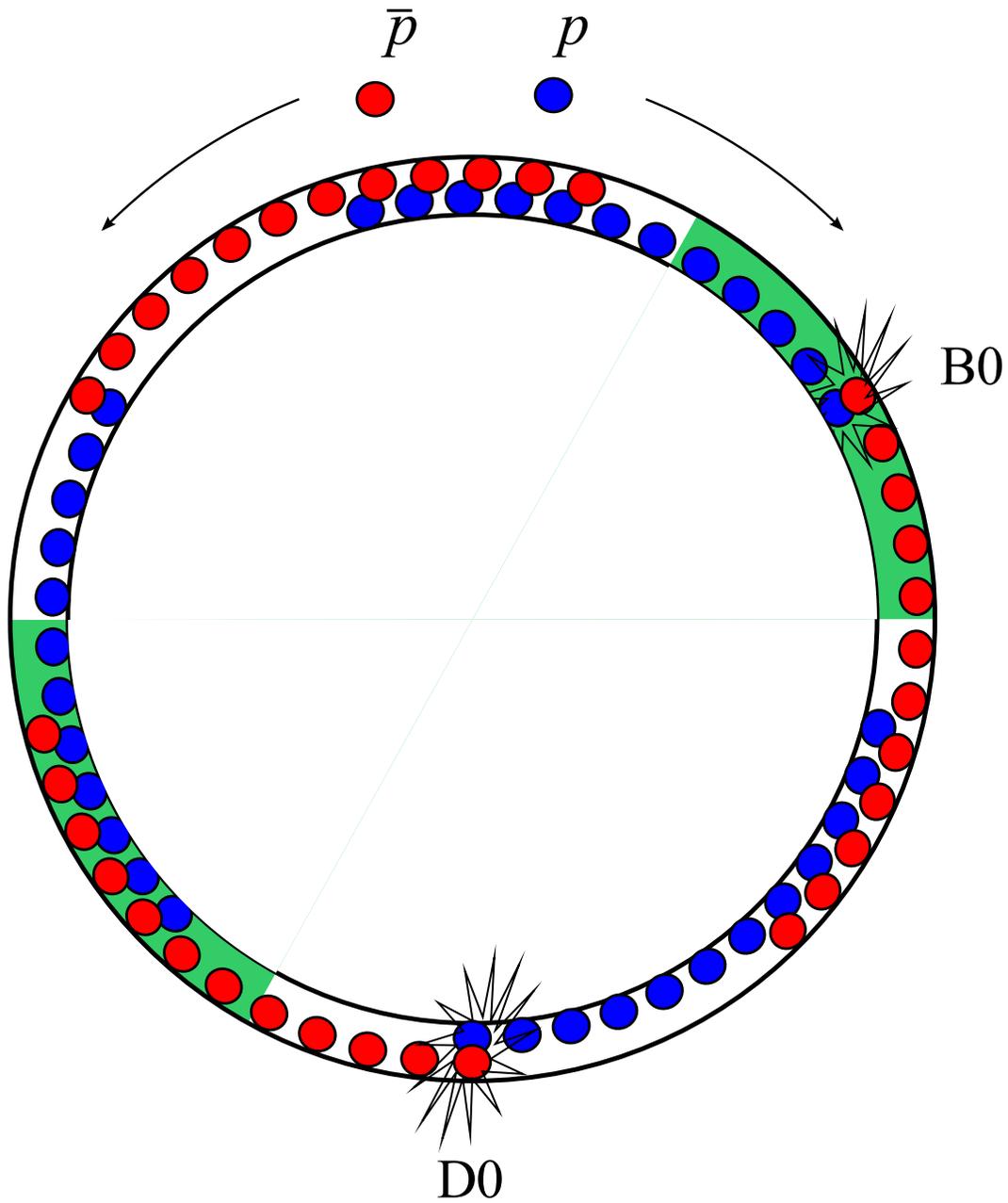}
\caption{\label{tev_fill_pattern}Schematic of the position of proton and antiproton bunches
in the Tevatron with 36 proton and 36 antiproton bunches.
The diagram shows the positions at a time when the lead bunch of the trains are
at the head-on collision location.
Head-on collisions occur at location~B0 and~D0.
The green shading indicates the part of the ring where beam-beam
collisions may occur in the simulations with six-on-six bunches.}
\end{figure}

Understanding the effect of unwanted long-range collisions among multiple
beam bunches in the 
design and operation of hadron colliders has received attention from other
authors \cite{pieloni,lifetrac} which underscores the importance for this
kind of simulation.
A schematic of the fill pattern of proton and
antiproton bunches in the Tevatron is shown in
Fig.~\ref{tev_fill_pattern}.
There are
three trains of twelve bunches for each species.
A train occupies approximately~$81.5^\circ$ separated by
a gap of about~$38.5^\circ$.
The bunch train and gap are replicated three times to fill the ring.
Bunches collide head-on at the B0 and D0 interaction points but undergo
long range (electromagnetic) beam-beam interactions at 136 other
locations around the ring\footnote{With three-fold symmetry of bunch
trains, train-on-train collisions occur at six
locations around the ring.
The collision of two trains of 12 bunches each results in bunch-bunch
collisions at 23 locations which when multiplied by six results in~138
collision points.
It is a straightforward computer exercise to enumerate these locations.
Two of these locations are distinguished as head-on while the remainder
are parasitic.\cite{shiltsev_beambeam}}.

Running the simulation with all~136 long-range IPs turns out to be very slow
so we only calculated beam-beam forces at the two main IPs and
and the long-range IPs
immediately upstream and downstream of them.
The transverse beta functions at the long-range collision
locations are much larger than the bunch length, so the beam-beam calculation
at those locations can be performed using only the 2D solver.

One interesting consequence of the fill pattern and the helical trajectory
is that any one of the 12 bunches in a train experiences collisions with
the~36 bunches in the other beam at different locations around the ring,
and in different transverse positions.
This results in a different tune and emittance growth for each bunch of
a train, but with the three-fold symmetry for the three trains.
In the simulation, emittance growth arises from the effects of impedance 
acting on bunches that have been perturbed by beam-beam forces.
The phenomenon of bunch dependent emittance growth is observed experimentally\cite{shiltsev_beambeam}.

\begin{figure}[htb]
\centering
\includegraphics*[width=\columnwidth]{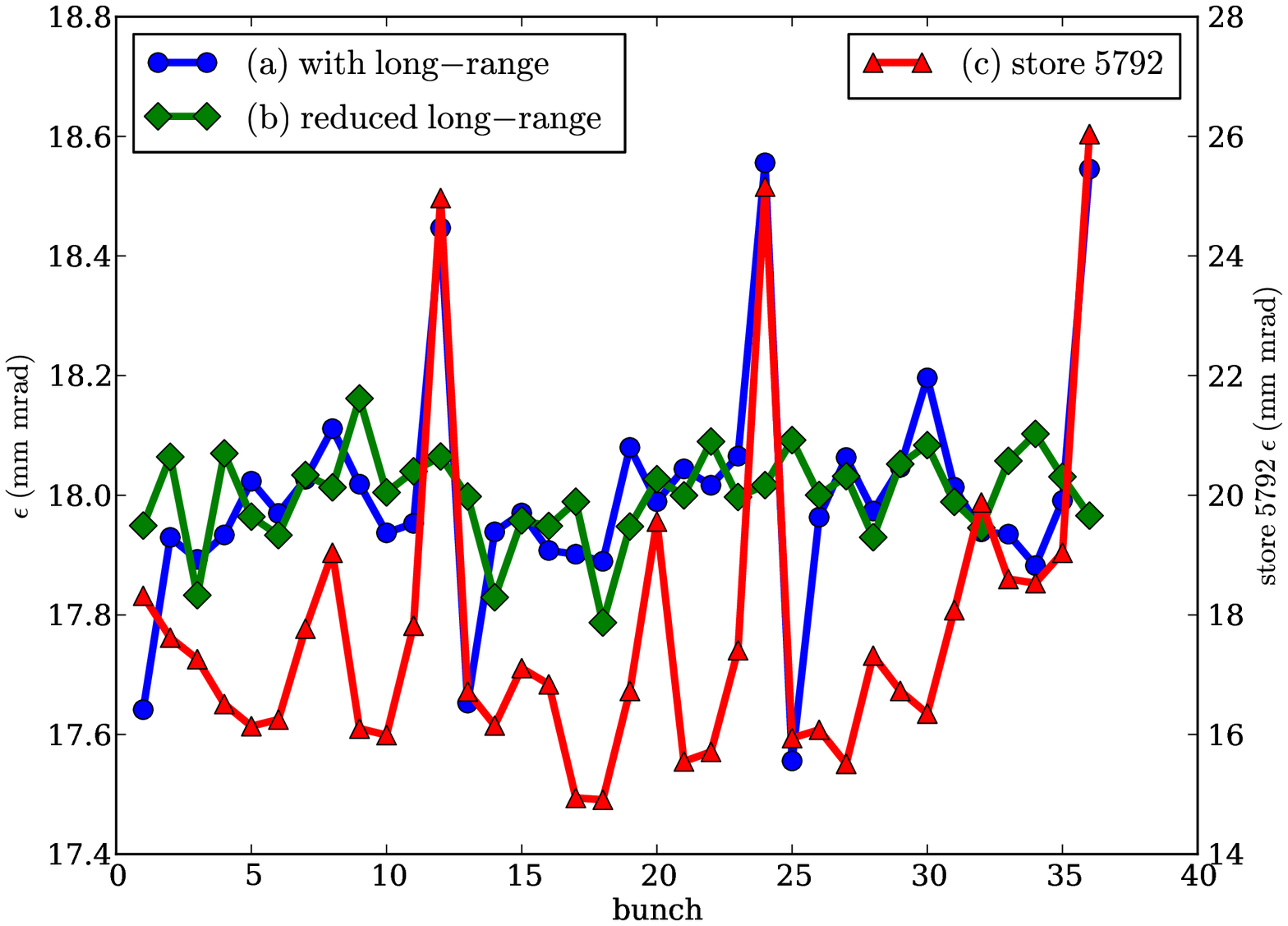}
\caption{The simulated and measured emittance of each Tevatron proton bunch
after running with 36 proton and 36 antiproton bunches.
Curves (a) and~(b) which show the emittance after 50000 simulated turns
are read with the left vertical axis.
Curve (a) results from a simulation with the nominal beam
spacing at the long-range IPs.
Curve (b) results from a simulation with the
hypothetical condition where the beam separation
at the long-range IPs is 100 times normal, suppressing the
effect of those long-range IPs.
Curve (c) is the measured emittance of bunches after 15 minutes of a particular
store (\#5792) of bunches in the Tevatron, and is read with the right
vertical axis.
}
\label{yemitgrowth}
\end{figure}

The beam-beam simulation with 36-on-36 bunches shows similar effects.
We ran a simulation of 36~proton on 36~antiproton bunches
for 50000 turns with the nominal helical orbit.
The proton bunches had $8.8 \times 10^{11}$ particles (roughly four
times the usual to enhance the effect) and the proton emittance was the typical $20 \pi \,\textrm{mm}\,\textrm{mrad}$.
The antiproton bunch intensity and emittance were both half the corresponding
proton bunch parameter.
The initial emittance for each proton bunch was the same so changes during
the simulation reflect the beam-beam effect.

Curve~(a) in Figure~\ref{yemitgrowth} shows the emittance for each
of the 36~proton bunches in a 36-on-36
simulation after 50000 turns of simulation.
The three-fold symmetry is evident.
The end bunches of the train (bunch 1, 13, 25) are clearly different from the interior bunches.
For comparison, curve~(c) shows the measured 
emittance taken during accelerator operations.
The observed bunch emittance variation is similar to the simulation results.
Another beam-beam simulation
with the beam separation at the closest head-on IP
expanded 100~times its nominal value resulted in curve~(b) of Figure~\ref{yemitgrowth}
 showing a much reduced bunch-to-bunch variation.
We conclude that the beam-beam effect at the long-range IPs is the origin
of the bunch variation observed in the running machine and
that our simulation of the beam helix is correct.

\section{\label{sec:TevAppSec}Tevatron applications}

\subsection{Single bunch features}

\begin{figure}
\includegraphics*[width=\columnwidth]{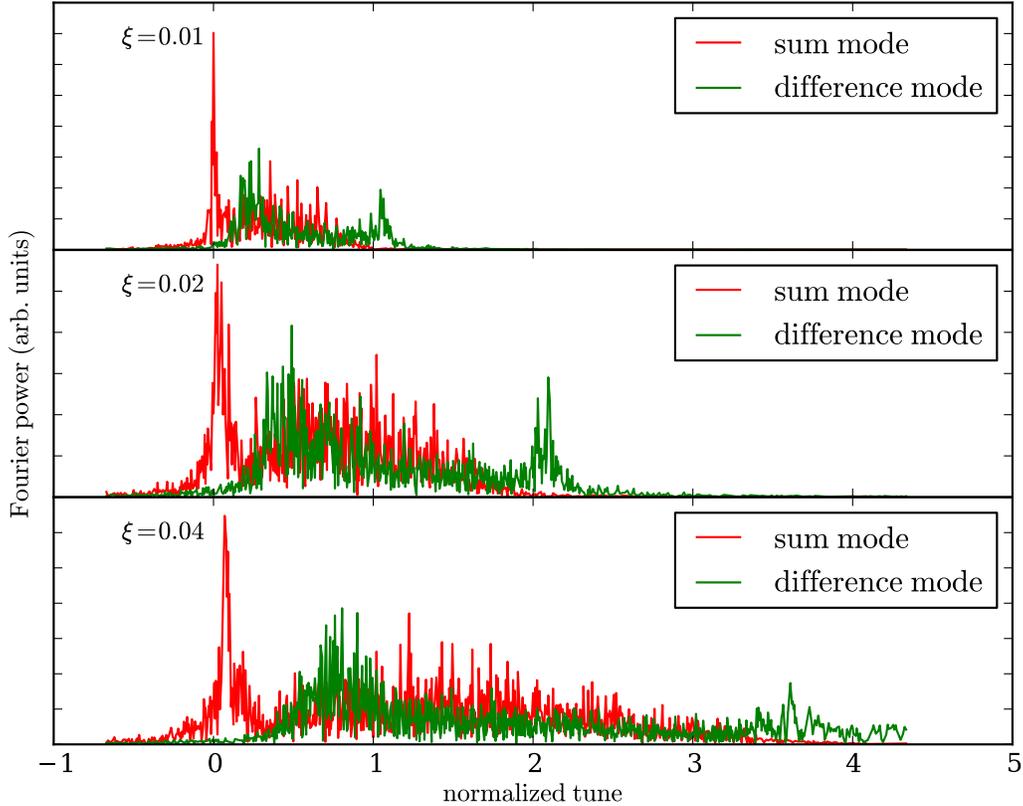}
\caption{Dipole mode spectra of the sum and difference offsets of
two beam centroids at three  beam
intensities corresponding to beam-beam parameter values for each beam
of~0.01, 0.02
and~0.04.
The vertical scale is in arbitrary units.}
\label{modes_2.2e11_8.8e11}
\end{figure}

We looked at the tune spectrum with increasing intensity for equal
intensity~$p$
and~$\bar p$ beams containing one bunch each.
As the intensity increases, the beam-beam parameter $\xi$ increases.
Fig.~\ref{modes_2.2e11_8.8e11} shows the spectrum of the sum and
difference of the two beam centroids for $\xi = 0.01, 0.02, 0.04$, corresponding
to beam bunches containing~$2.2\times 10^{11}$, $4.4 \times 10^{11}$
and~$8.8\times 10^{11}$ protons.
The abscissa is shifted so the base tune is at~0 and normalized in units of
the beam-beam parameter at a beam intensity of~$2.2\times 10^{11}$.
The coherent~$\sigma$ and~$\pi$ mode peaks are expected to be present
in
the spectra of the sum and difference signals of the two beam centroids.
The coherent $\sigma$ modes are evident at~0, while
the coherent $\pi$ modes should slightly greater than~1, 2, and~4
respectively.
Increasing intensity also causes larger induced wake fields
which broaden the mode peaks, especially the~$\pi$ mode, as shown in
Fig.~\ref{modes_2.2e11_8.8e11}.

\begin{figure}
\includegraphics*[width=\columnwidth]{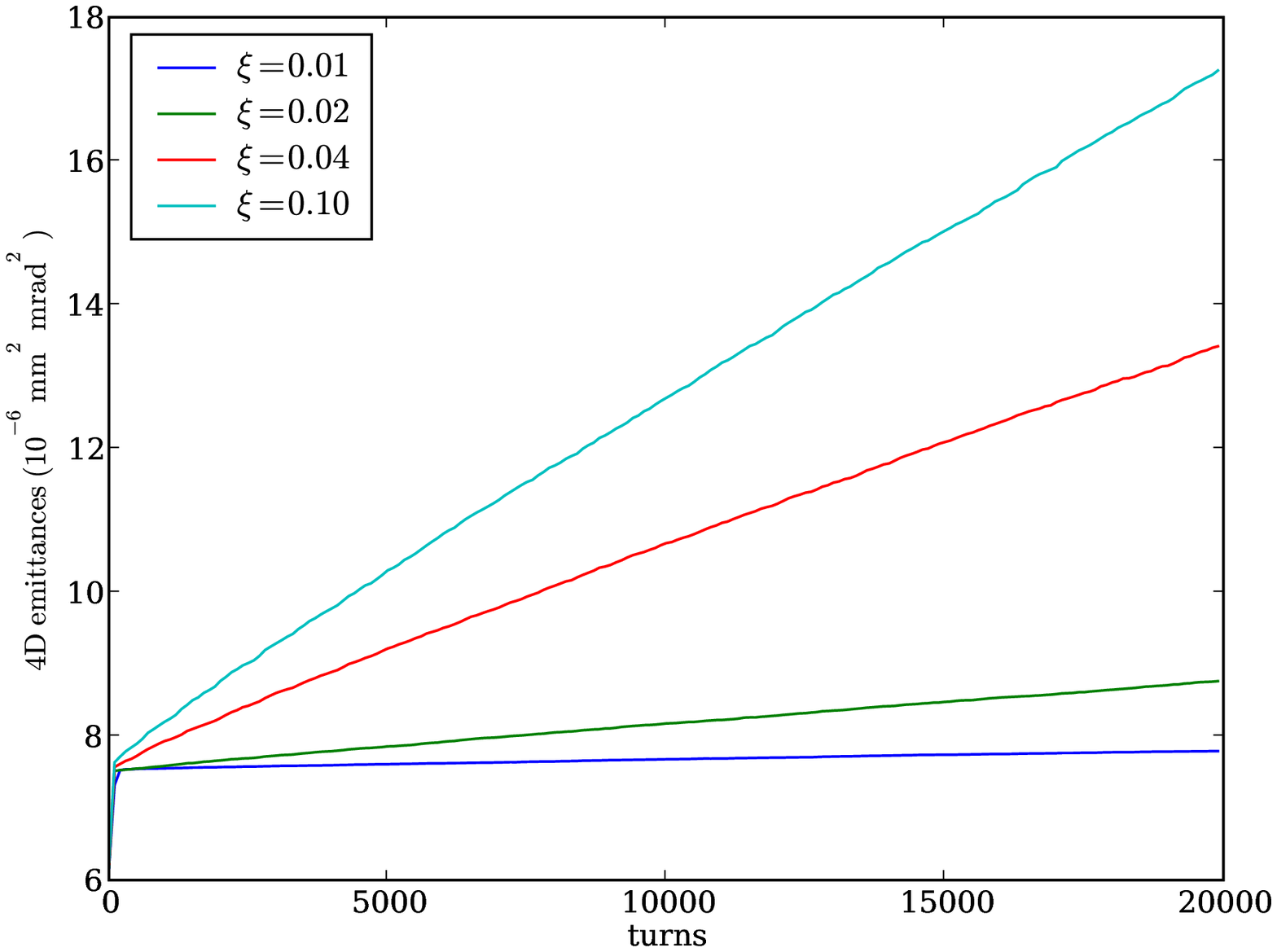}
\caption{The evolution of 4D emittances for beam-beam parameters
of~0.01, 0.02, and~0.04 which correspond to
intensities of (a)~$2.2 \times 10^{11}$, (b)~$4.4 \times 10^{11}$, (c)~$8.8 \times 10^{11}$, and (d)~$1.1 \times 10^{12}$ protons  per bunch.}
\label{emittances_2.2e11_1.1e12}
\end{figure}

\begin{figure}
\includegraphics*[width=\columnwidth]{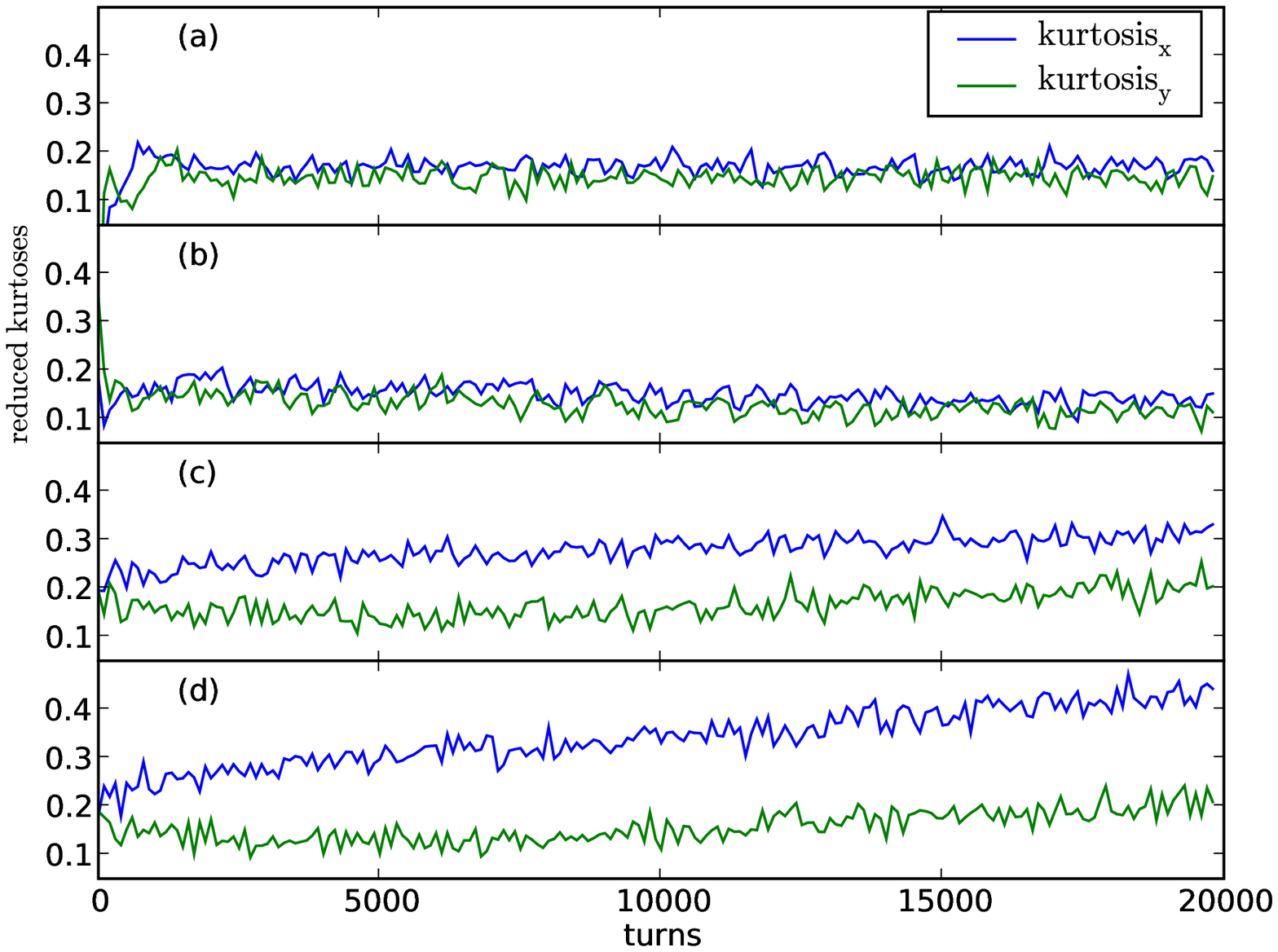}
\caption{The evolution of (reduced) kurtosis of the particle distribution
for intensities of (a)~$2.2 \times 10^{11}$, (b)~$4.4 \times 10^{11}$, (c)~$8.8 \times 10^{11}$, and (d)~$1.1 \times 10^{12}$ protons per bunch.
}
\label{kurtoses_2.2e11_1.1e12}
\end{figure}

The 4D emittances at higher intensities show significant growth over 20000
turns as shown in Fig.~\ref{emittances_2.2e11_1.1e12}.
The kurtosis excess of
the two beams remains slightly positive for the nominal intensity, but shows
a slow increase at higher intensities indicating the the beam core is being
concentrated as shown in~Fig.~\ref{kurtoses_2.2e11_1.1e12}.
Concentration of the bunch core while emittance is growing indicates
the development of filamentation and halo.

\subsection{Simulation of bunch length, synchrotron motion and beam-beam interactions}
\begin{figure*}[tb]
\includegraphics*[width=\textwidth]{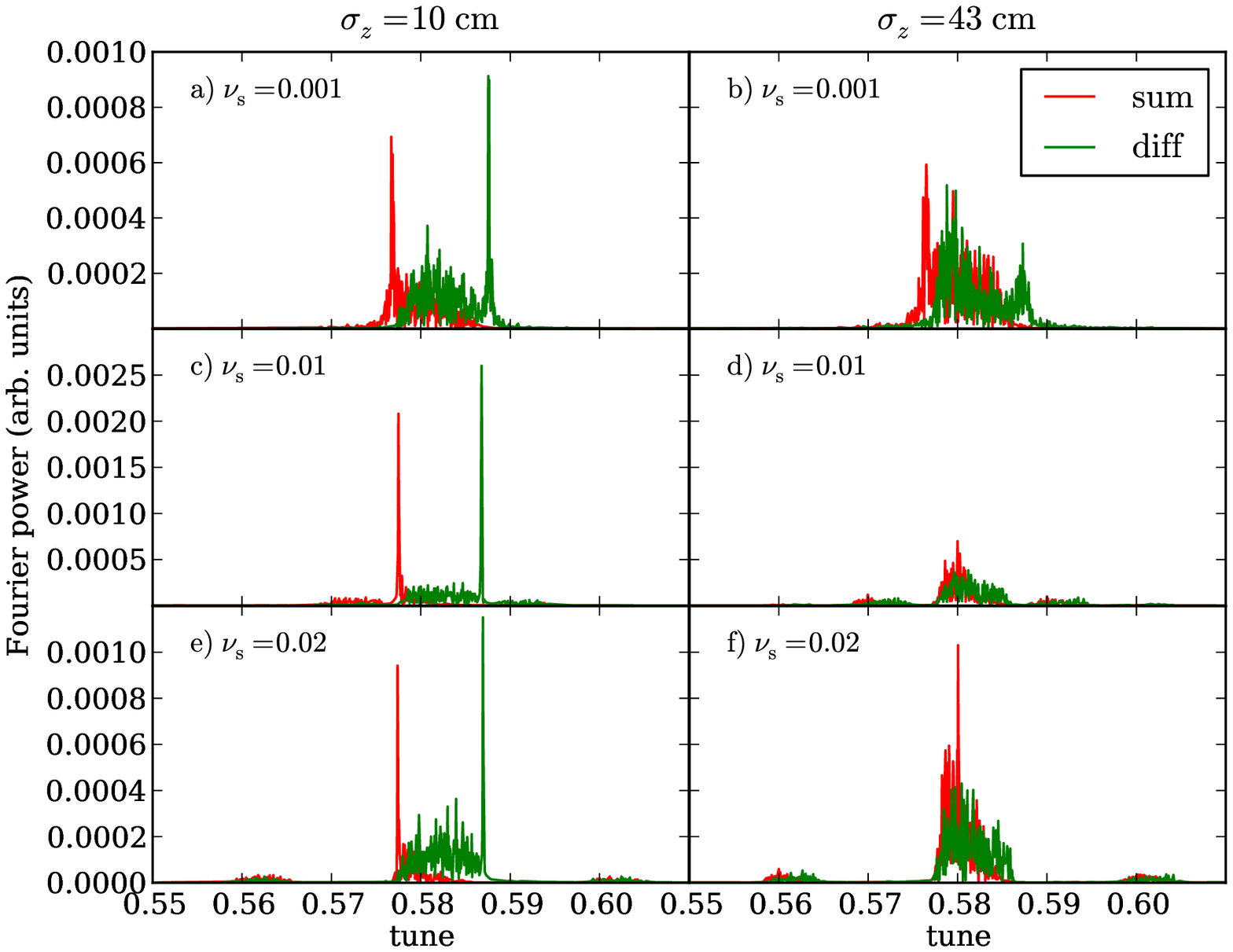}
\caption{\label{qs_scan}Simulated one-on-one bunch $y$ plane
$\sigma$ and~$\pi$ mode tune spectra for short bunches (a, c, e)
and long bunches (b, d, f), for three different synchtrotron tunes, with
a Tevatron-like lattice.}
\end{figure*}

Synchrotron motion in extended length bunches modifies the effects of the
beam-beam interaction by shifting and suppressing the coherent modes.
The plots in Fig.~\ref{qs_scan} show simulated spectra for sum and difference
signals of the beam centroid offsets for
one-on-one bunch collisions in a ring with Tevatron-like
optics, with both short and long bunches, at three different synchrotron tunes.
The sum signal will contain the $\sigma$~mode while the difference
signal will contain the $\pi$~mode.
In this Tevatron simulation, the beam strength is set so that the beam-beam parameter is~0.01, the base tune in the vertical plane is~0.576,
and $\beta_y$ is approximately~$30\,{\rm cm}$.
Subplots~{\textit a} and~{\textit b} of Fig.~\ref{qs_scan} show that with small synchrotron tune both the~$\sigma$
and~$\pi$ mode peaks are evident with~short and~long bunches.
The $\sigma$ mode peak is at the proper place, with the $\pi$ mode peak
shifted upwards by the expected amount, but with
longer bunches (subplots {\textit c} and~{\textit d}) the incoherent continuum is enhanced and the strength of the
coherent peaks is reduced.
When the synchrotron tune is the same as or larger than the beam-beam splitting (subplots {\textit e} and {\textit f}), short bunches
still exhibit strong coherent modes, but with long bunches the coherent
modes are significantly diluted.
In the case of long bunches, the $\sigma$~mode has been shifted upwards to
0.580, and the $\pi$~mode
is not clearly distinguishable from the continuum.
At $\nu_s$ of~$0.01$ and~$0.02$, the
synchrobetatron side bands are clearly evident.

\subsection{Multi-bunch mode studies}
When the Tevatron is running in its usual mode, each circulating beam 
contains 36~bunches.
Every bunch in one beam interacts with every bunch in the opposite beam,
though only two interaction points
are useful for high energy physics running.
The other 136 interaction points are unwanted and detrimental to
beam lifetime and luminosity.
The beam orbit is deflected in a helical shape by electrostatic separators
to reduce the impact of these unwanted collisions,
so the beams are transversely separated from each other in
all but the two high-energy
physics interaction points.
Because of the helical orbit, the beam separation is different at each parasitic collision location.
For instance, a bunch near the front of the bunch train will undergo
more long-range close the the head-on interaction point, compared to a bunch
near the rear of the bunch train.
A particular bunch experiences collisions at specific interaction points
with other bunches each of which has its own history
of collisions.
This causes bunch-to-bunch variation in disruption
and emittance growth as will be demonstrated below.

We will begin the validation and exploration of the multi-bunch implementation
starting with runs of two-on-two bunches and six-on-six bunches before moving
on to investigate the situation with the full Tevatron bunch fill
of 36-on-36 bunches.
Two-on-two bunches will demonstrate the bunches coupling amongst each other,
but will not be enough to demonstrate the end bunch versus interior bunch
behavior that characterizes the Tevatron.
For that, we will look at
six-on-six bunch runs.

In these studies, we are only filling the ring with at most six~bunches
in a beam.
Referring to Fig.~\ref{tev_fill_pattern}, we see that only the head-on
location at~B0 is within
the green shaded region where beam-beam collisions may occur with
six bunches in each beam.
Because of the beam-beam collisions, each bunch is weakly coupled to
every other bunch which gives rise to multi-bunch collective modes.

\begin{figure}[tb]
\includegraphics*[width=\columnwidth]{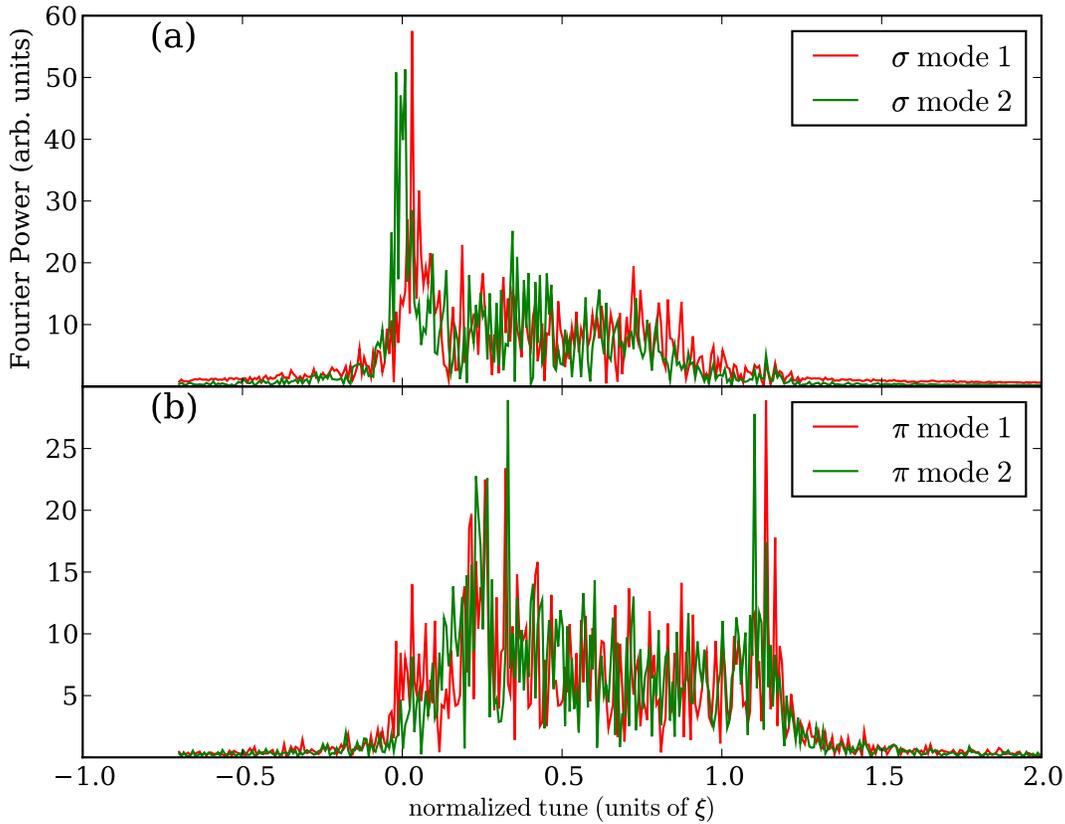}
\caption{\label{two-on-two}Mode tune spectrum for a two on two bunch run
at~$2.2\times 10^{11}$ particles/bunch ($\xi = 0.01$).
Figure (a) shows the two
modes that are most like~$\sigma$ modes.
$\sigma$ mode~1 is $0.53 y_1 + 0.53 y_2 +  0.59 y_3 - 0.31 y_4$,
$\sigma$ mode~2 is $0.39 y_1 + 0.49 y_2 - 0.46 y_3 - 0.63 y_4$.
  Figure (b) shows the
two $\pi$-like modes.
$\pi$ mode~1 is $0.74 y_1 - 0.66 y_2 -  0.08 y_3$,
$\pi$ mode~2 is $0.12 y_1 + 0.20 y_2 - 0.66 y_3 + 0.31 y_4$.
The absolute scale of the Fourier power is arbitrary, but the relative
scales of plots~(a) and~(b) are the same.
}
\end{figure}

We began the investigation of these effects with a
simulation of beams with two bunches each.
The bunches are separated by 21~RF buckets as they are are in normal Tevatron
operations.
Collisions occur at the head-on location and at
parasitic locations 10.5~RF buckets distant on either side of the head-on location.
To make any excited modes visible, we ran with~$2.2 \times 10^{11}$ particles,
which gives a single bunch beam-beam parameter of~$0.01$.
There are four~bunches in this problem.
We label bunch~1 and~3 in beam 1 (proton) and bunch~2 and~4 in beam 2 (antiproton)
with mean $y$~positions of the bunches $y_1,\ldots y_4$.
By diagonalizing the covariance matrix of the turn-by-turn bunch
centroid deviations, we determine four modes, shown in Fig.~\ref{two-on-two}.
Fig.~\ref{two-on-two}(a) shows the splitting of the~$\sigma$ mode.
The coefficients of the two modes indicate that this mode is primarily
composed of the sum of  corresponding beam bunches (1 with 2, 3 with 4)
similar to the $\sigma$ mode in
the one-on-one bunch case.
The other two modes in Fig~\ref{two-on-two}(b) have the character and location
in tune space of the
$\pi$~mode, from their coefficients and also their reduced strength compared
to the~$\sigma$ mode.

\begin{figure*}[tb]
\begin{tabular}{@{}cc}
\includegraphics*[width=0.5\textwidth]{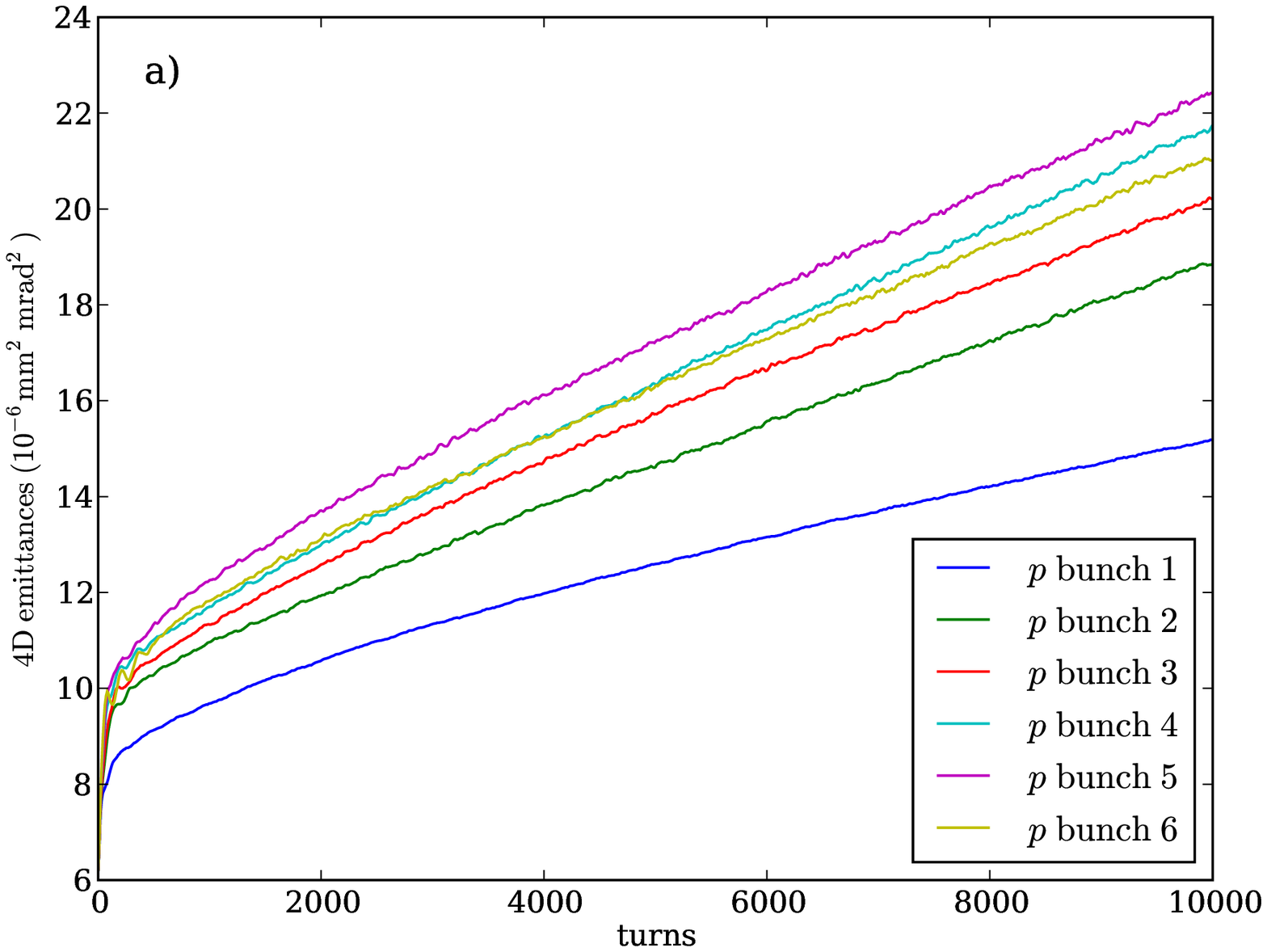} &
\includegraphics*[width=0.5\textwidth]{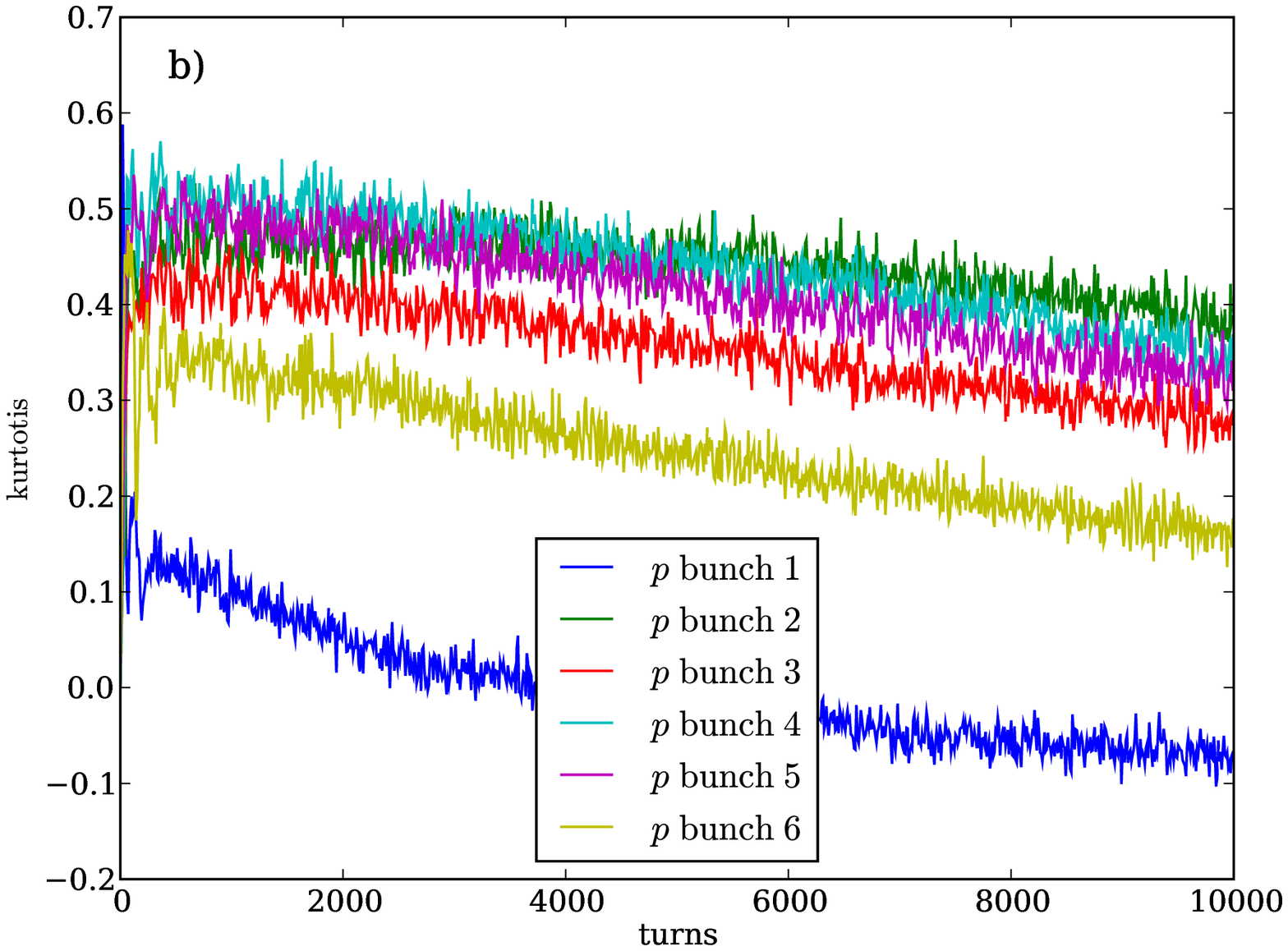}
\end{tabular}
\caption{\label{six-on-six-nobump}A six-on-six bunch Tevatron run with
$8.8\times 10^{11}$ particles/bunch: (a) The turn-by-turn evolution of 4D emittance of each of the six bunches.
 (b) The turn-by-turn evolution of $y$~kurtosis of the six bunches.
}
\end{figure*}

With six on six bunches, features emerge that are clearly bunch position
specific.
Fig.~\ref{six-on-six-nobump}(a) shows the turn-by-turn evolution
of 4D emittance and (b) $y$~kurtosis for each of the
six proton bunches.
It is striking that bunch~1, the first bunch in the sequence, has a lower
emittance growth than all the other bunches.
Emittance growth increases faster with increasing bunch number from
bunches 2--5, but bunch~6 has a lower emittance growth than even bunch~4.
The kurtosis of bunch 1 changes much less than that of any of the other bunches,
but bunches 2--5 have a very similar evolution, while bunch~6 is markedly closer
to bunch~1.
One difference between the outside bunches (1 and~6) and the inside bunches
(2--5) is that they have only one beam-beam interaction at the parasitic
IP closest to the head-on collision, while the inside bunches have one
collision before the head-on IP, and one after it.
The two parasitic collision points closest to the head-on collision point
have the smallest separation of any of the parasitics, so interaction there
would be expected to disrupt the beam more than interactions at other
parasitic locations.

\begin{figure*}[tb]
\begin{tabular}{@{}cc}
\includegraphics*[width=0.5\textwidth]{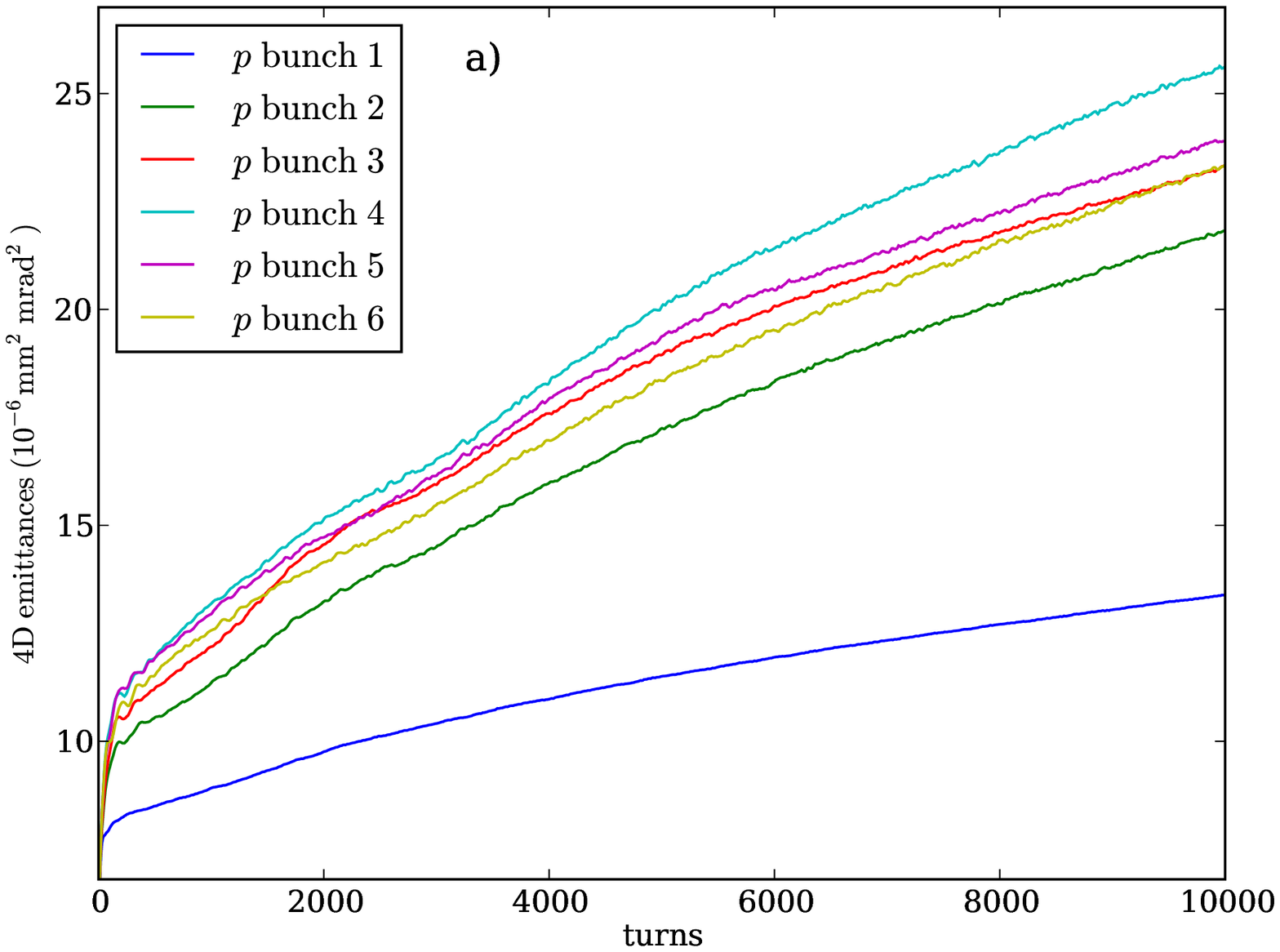}
&
\includegraphics*[width=0.5\textwidth]{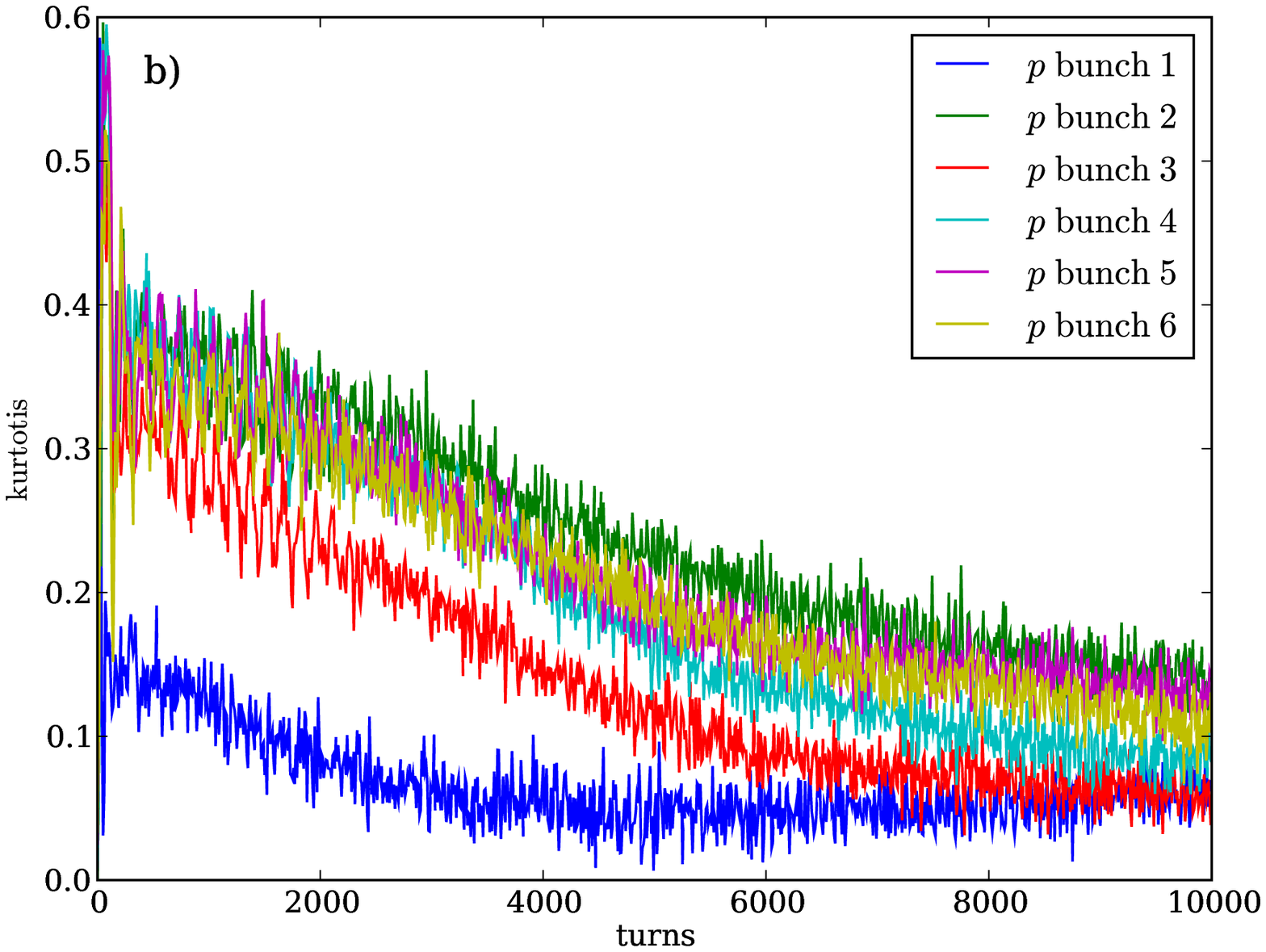}
\end{tabular}
\caption{\label{six-on-six-bump}In a six-on-six bunch Tevatron run with
$8.8 \times 10^{11}$ particles/bunch, with the beam spacing at the first
parasitic IP downstream of the head-on location artificially
increased: (a) The 4D emittance of each of the six bunches as a function
of~turn.  (b) the $y$~kurtosis of the six bunches as a function of~turn.
}
\end{figure*}

To test this hypothosis, we did two additional runs.
In the first, the beam separation at the  parasitic IP immediately downstream
of the head-on IP was
artificially increased in the simulation  so as to have essentially no effect.
The effect of this is that the first proton bunch will not have any
beam-beam collisions at an IP close to the head-on IP, while all the
other bunches will have one collision at a near-head-on IP.
The corresponding plots of emittance and kurtosis are shown in 
Fig.~\ref{six-on-six-bump}.
The kurtosis data shows that bunches 2--5 which all suffer one beam-beam
collision at a close parasitic IP are all together while bunch~1  which
does not have a close IP collision is separated from the others.

\begin{figure*}[tb]
\begin{tabular}{@{}cc}
\includegraphics*[width=0.5\textwidth]{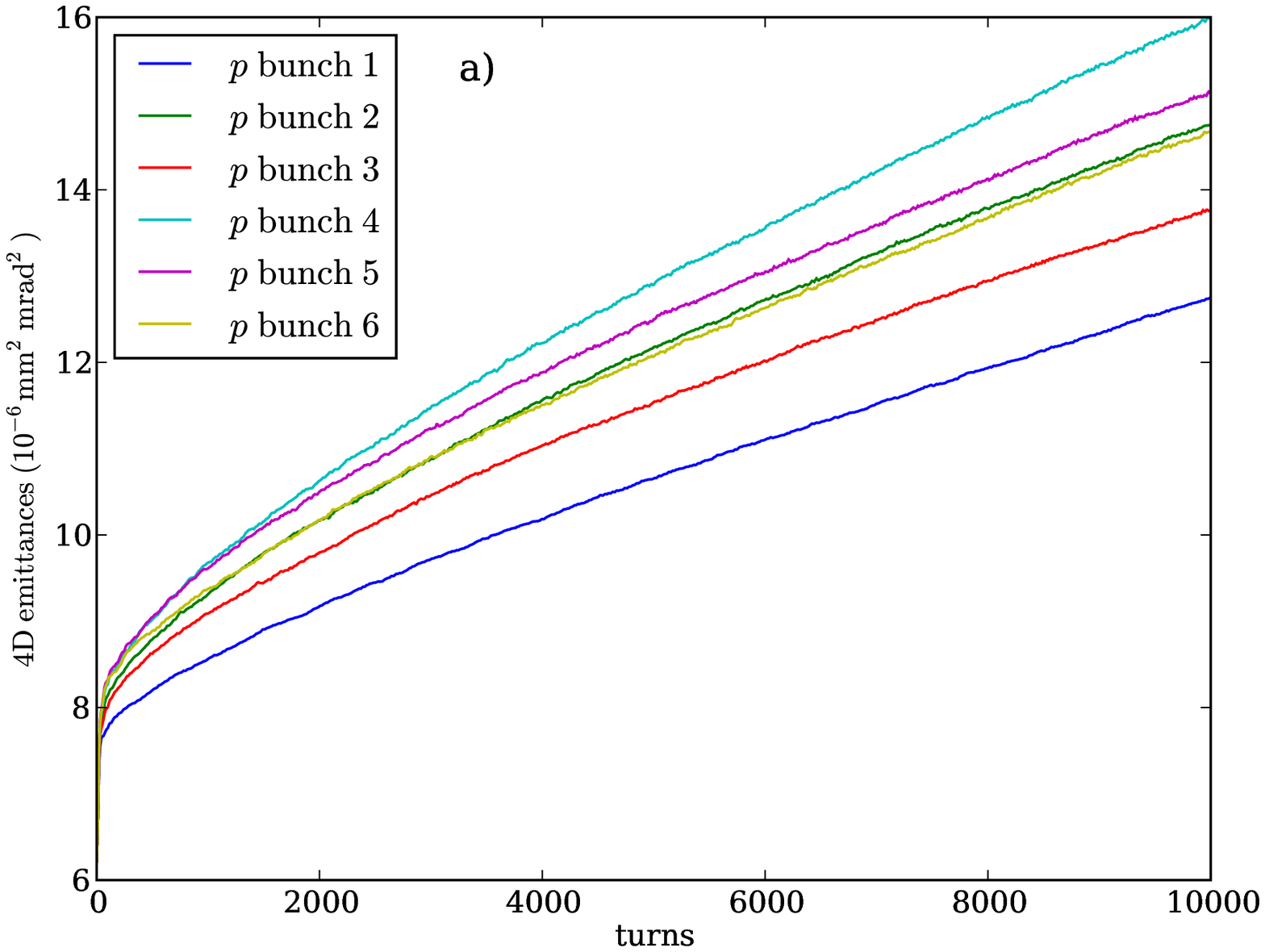}
&
\includegraphics*[width=0.5\textwidth]{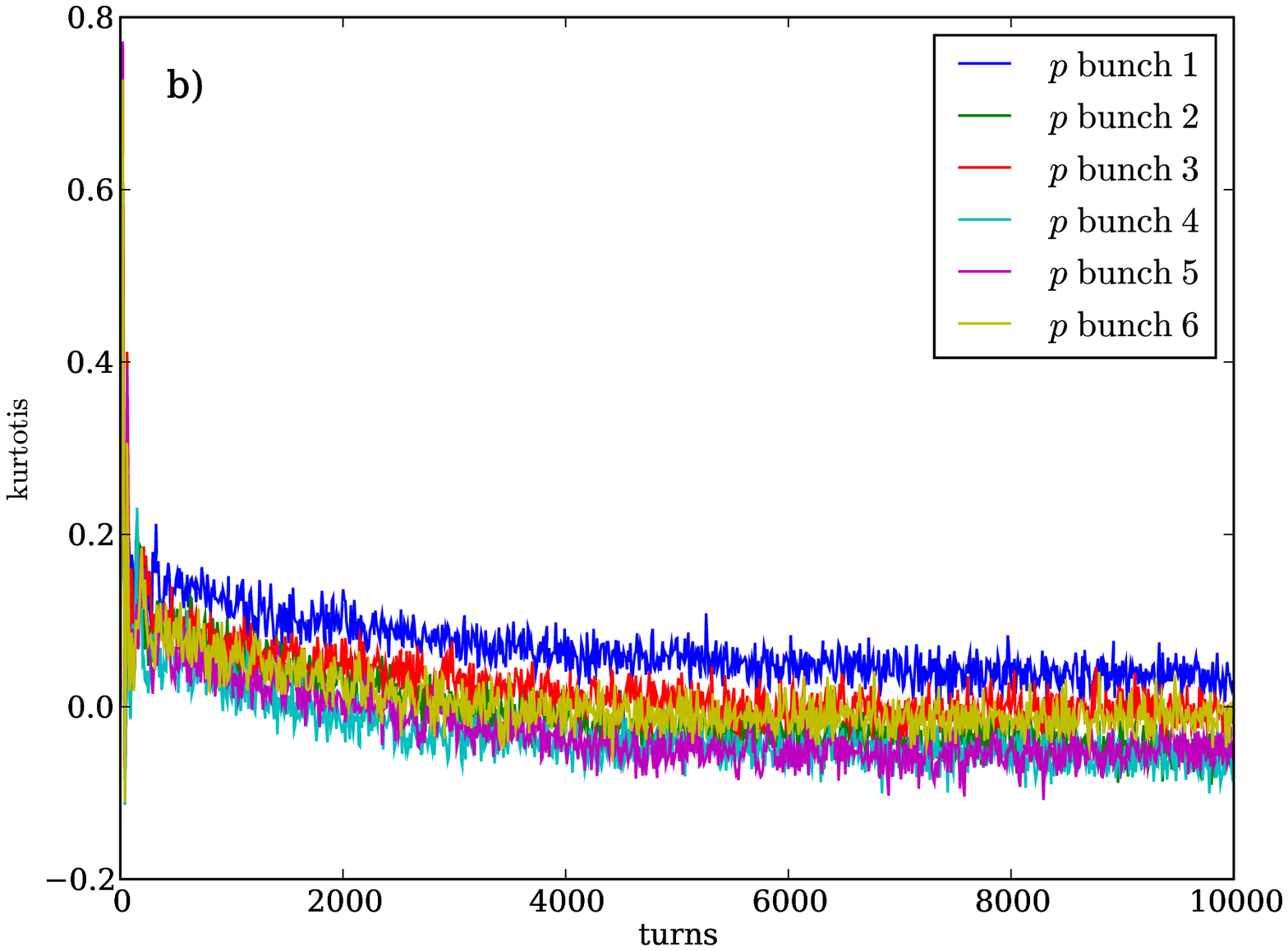}
\end{tabular}
\caption{\label{six-on-six-bump2}In a six-on-six bunch Tevatron run with
$8.8\times 10^{11}$ particles/bunch, with the both nearest upstream
and downstream parasitic IP artificially
widened: (a) The 4D emittance of each of the six bunches as a function
of~turn.  (b) the $y$~kurtosis of the six bunches as a function of~turn.
}
\end{figure*}

Emittance and kurtosis growth in simulations where the beam separation
at the closest upstream and downstream
parasitic IPs was increased is shown in
Fig.~\ref{six-on-six-bump2}.
In this configuration no bunch suffers a strong beam-beam collision at a parasitic
IP close to the head-on location so the kurtosis of all the bunches evolves
similarly.

\section{\label{sec:LowerChrom}Lower chromaticity threshold}

During the Tevatron operation in 2009 the limit for
increasing the initial luminosity was determined by particle losses
in the so-called squeeze phase \cite{AVPAC09}. At this stage the
beams are separated in the main interaction points (not colliding head-on),
and the machine optics is gradually changed to decrease the beta-function
at these locations from 1.5 m to 0.28 m.

With proton bunch intensities currently approaching
$3.2 \times 10^{11}$ particles, the chromaticity of the Tevatron has to
be managed carefully to avoid the development of a  head-tail instability.
It was determined experimentally that after the head-on collisions are initiated,
the Landau damping introduced by beam-beam interaction is strong enough to
maintain beam stability at chromaticity of +2 units (in Tevatron
operations, chromaticity is $\Delta \nu / ({\Delta p}/p)$.)
At the earlier stages of the collider cycle, when beam-beam effects are limited
to long-range interactions the chromaticity was kept as high
as 15 units since the concern was that the Landau damping is insufficient
to suppress the instability. At the same time, high chromaticity causes particle
losses which are often large enough to quench the superconducting magnets, 
and hence it is desireable to keep it at a reasonable minimum.

\begin{table}[hbt]
\caption{\label{beam_parameters}
Beam parameters for Tevatron simulation}
\begin{ruledtabular}
\begin{tabular}{lr}
\textbf{Parameter} & value \\
\hline
beam energy & $980\,\mbox{GeV}$ \\
$p$ particles/bunch & $3.0 \times 10^{11}$  \\
$\bar p$ particles/bunch & $0.9 \times 10 ^{11}$ \\
$p$ tune $(\nu_x,\nu_y)$ & (20.585,20.587) \\
$p$ (normalized) emittance & $20 \pi\,\textrm{mm}\,\textrm{mrad} $ \\
$\bar p$ tune $(\nu_x,\nu_y)$ & (20.577,20.570) \\
$\bar p$ (normalized) emittance & $6 \pi\,\textrm{mm}\,\textrm{mrad}$ \\
synchrotron tune $\nu_s$ & 0.0007 \\
slip factor & 0.002483 \\
bunch length (rms) & $43\,{\rm cm}$ \\
$\delta p/p$ momentum spread & $1.2 \times 10 ^{-4}$ \\
effective pipe radius & $3\,{\rm cm}$ \\
\end{tabular}
\end{ruledtabular}
\end{table}

\begin{figure}[htb]
\centering
\includegraphics*[width=\columnwidth]{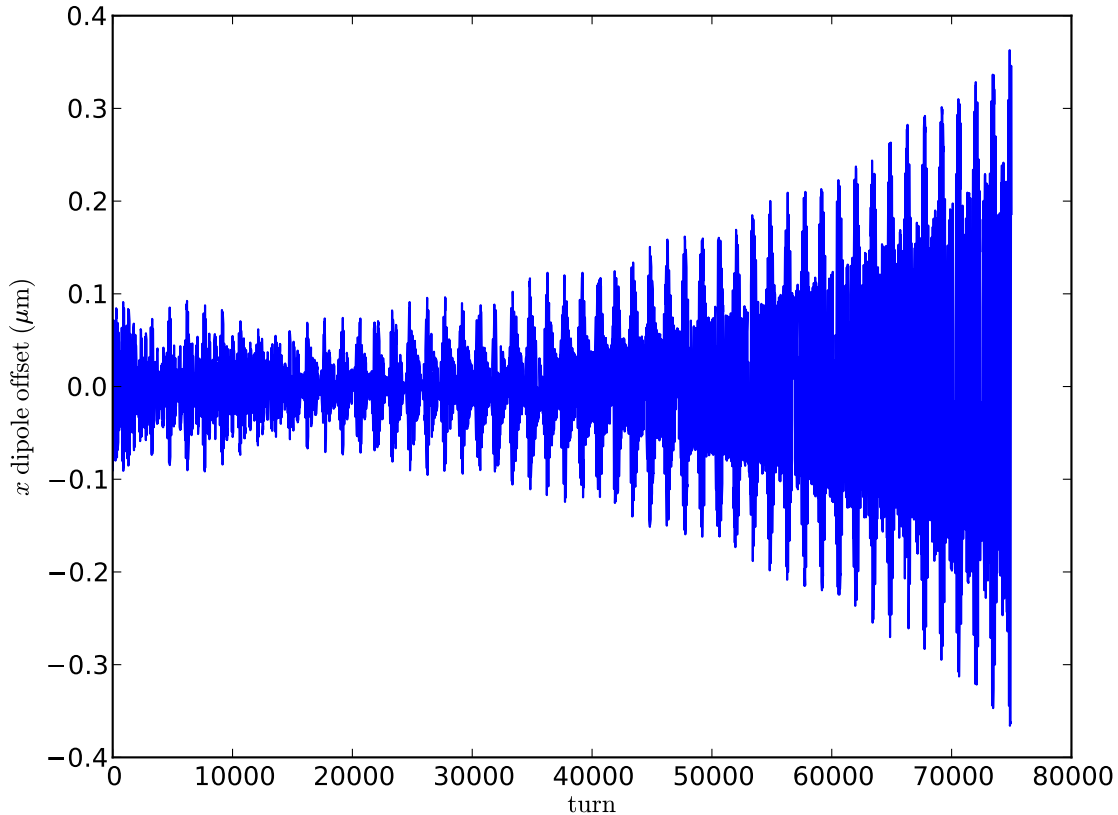}
\caption{The x dipole moment in a simulation with $C = -2$ and no beam-beam effect showing the development of instability.}
\label{x_dipole_nobb}
\end{figure}

\begin{figure}[htb]
\centering
\includegraphics*[width=\columnwidth]{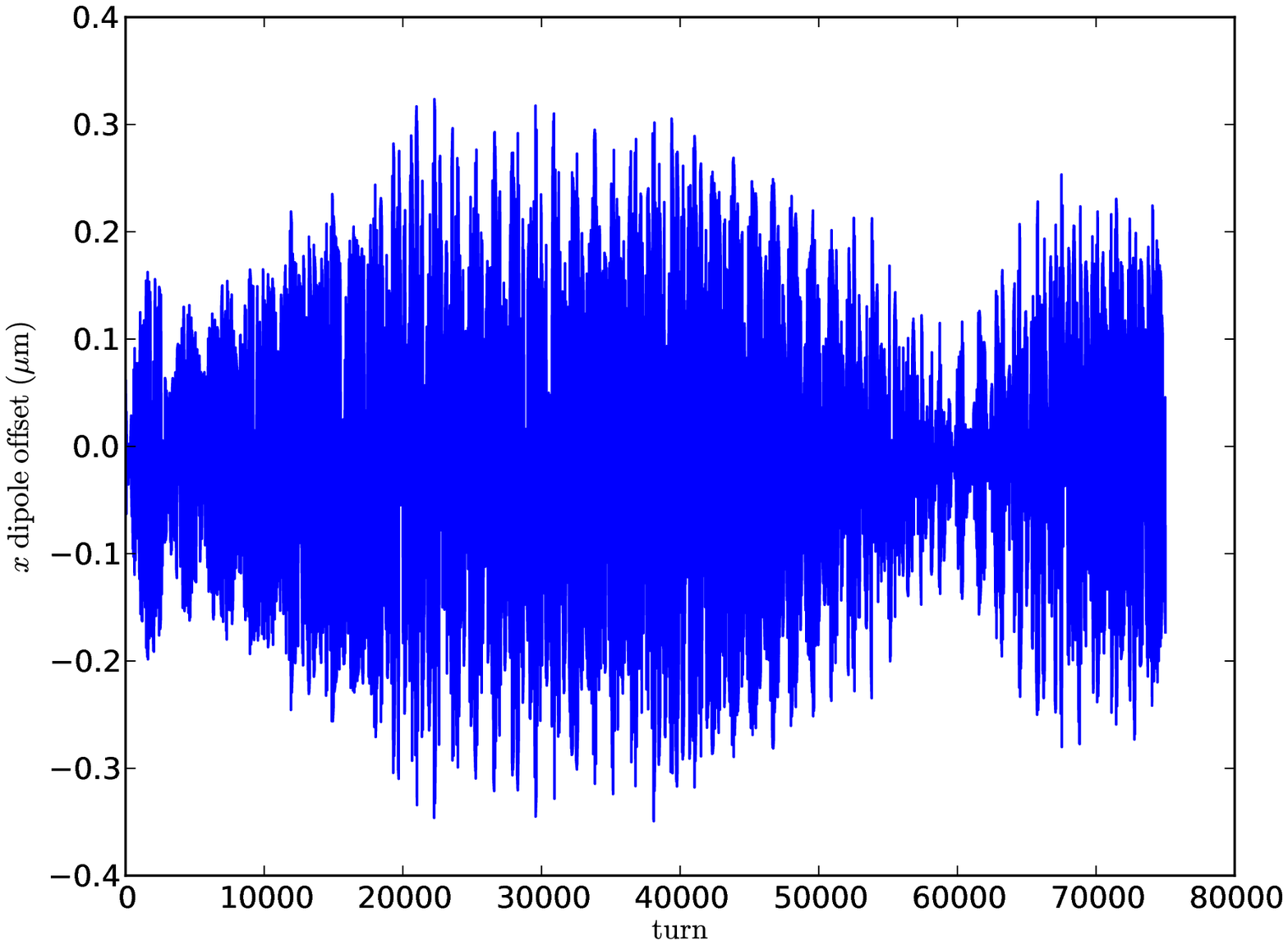}
\caption{The x dipole moment of a representative bunch in a 36-on-36 simulation with $C = -2$ with beam-beam effects and beams separated showing no obvious instability within the
limits of the simulation.}
\label{x_dipole_chrm_m2}
\end{figure}

\begin{figure}[htb]
\centering
\includegraphics*[width=\columnwidth]{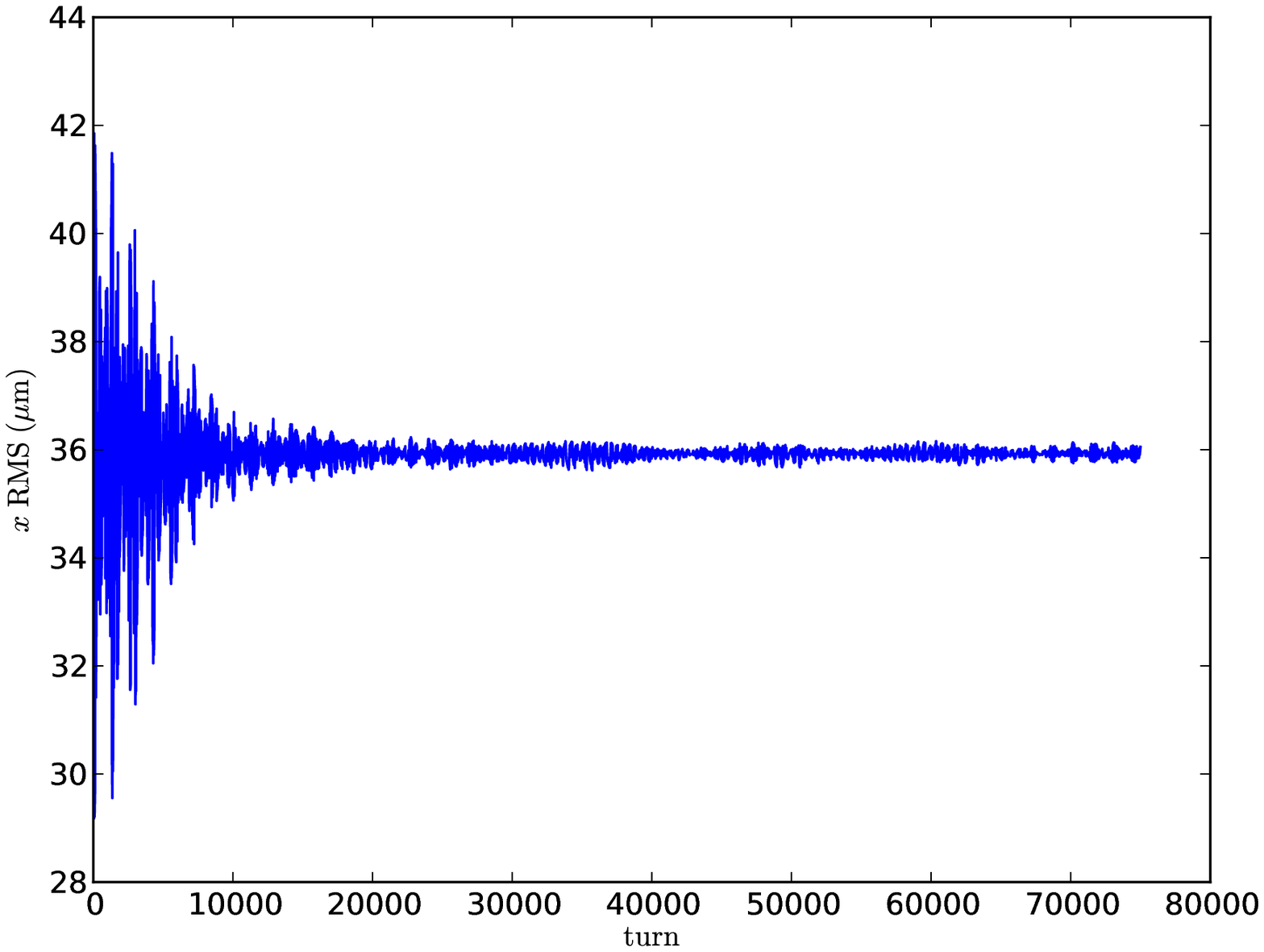}
\caption{The x RMS moment of a representative bunch in a 36-on-36 simulation with $C = -2$ with beam-beam effects and beams separated showing no obvious instability within the
limits of the simulation.}
\label{x_rms_chrm_m2}
\end{figure}

Our multi-physics simulation was used to determine the safe lower limit for 
chromaticity.
The simulations were performed with starting beam parameters listed in
Table~\ref{beam_parameters}.
With chromaticity set to -2 units, and no beam-beam effect, the
beams are clearly unstable as seen in Fig.~\ref{x_dipole_nobb}.
With beams separated, turning on the beam-beam effect prevents rapid
oscillation growth during the simulation as shown
in~Fig.~\ref{x_dipole_chrm_m2}.
The bursts of
increased amplitude is sometimes indicative of the onset of instability,
but it is not obvious within the limited duration of this run.
The RMS size of the beam also does not exhibit any 
obvious unstable tendencies as shown in Fig.~\ref{x_rms_chrm_m2}.

Based on these findings the chromaticity in the squeeze was lowered by a factor of
two, and presently is kept at 8-9 units. This resulted in a significant decrease
of the observed particle loss rates (see, {\textit e.g.}, Fig.~5 in \cite{AVPAC09}).

\section{Summary}

The key features of the developed simulation include fully three-dimentional 
strong-strong multi-bunch 
beam-beam interactions with multiple interaction points, transverse resistive 
wall impedance, and chromaticity. The beam-beam interaction model
has been shown
to reproduce the location and evolution of synchrobetatron modes
characteristic of the  3D strong-strong beam-beam interaction
observed in experimental data from the VEPP-2M collider.
The impedance calculation with macroparticles excites both the strong and
weak head-tail instabilities with thresholds and growth rates
that are consistent with expectations from
a simple two-particle model and Vlasov calculation.
Simulation of the interplay between the helical beam-orbit, long range
beam-beam interactions
and the collision pattern qualitatively matches observed patterns of
emittance growth.

The new program is a valuable tool for evaluation of the interplay between the
beam-beam effects and transverse collective instabilities.
Simulations have been successfully used to support the change of chromaticity
at the Tevatron, demonstrating that even the reduced beam-beam effect from
long-range collisions may provide enough Landau damping to
prevent the development of head-tail instability.
These results were used in Tevatron operations
to support a change of chromaticity during the transition to collider mode
optics, leading to a factor of two decrease in proton losses, and thus improved
reliability of collider operations.

\begin{acknowledgments}
We thank J. Qiang and R. Ryne of~LBNL for the use of and assistance with the BeamBeam3d program.
We are indebted to V.~Lebedev and Yu.~Alexahin for useful discussions.
This work was supported by the United States Department of Energy under contract DE-AC02-07CH11359 and the ComPASS project funded through the Scientific Discovery through Advanced Computing program in the DOE Office of High Energy Physics.
This research used resources of the National Energy Research Scientific Computing Center, which is supported by the Office of Science of the U.S. Department of Energy under Contract No. DE-AC02-05CH11231.
This research used resources of the Argonne Leadership Computing Facility at Argonne National Laboratory, which is supported by the Office of Science of the U.S. Department of Energy under contract DE-AC02-06CH11357.
\end{acknowledgments}










\end{document}